\documentclass{PoS}

\usepackage{graphicx}
\usepackage{amsmath,amssymb}
\usepackage{psfrag}
\newcommand{\ba}{\begin{eqnarray}}
\newcommand{\ea}{\end{eqnarray}}
\newcommand{\be} {\begin{equation}}
\newcommand{\ee} {\end{equation}}

\newcommand{\order}{O}

\title{Heavy flavour phenomenology from lattice QCD}

\ShortTitle{Heavy flavour phenomenology from lattice QCD}

\author{\speaker{Elvira G\'amiz}\\
        Department of Physics, University of Illinois, Urbana, IL 61801, USA\\
        E-mail: \email{megamiz@illinois.edu} 
        }

\abstract{The focus of this report is  the lattice calculation of 
hadronic parameters relevant to heavy flavour phenomenology. 
I will review recent results and the  current status of studies of the $B$ 
and $D$ decay constants, semileptonic decay form factors, 
$B^0-\bar B^0$ mixing, and determinations of the quark masses $m_c$ 
and $m_b$. Some studies of heavy flavour observables
have used current lattice results to derive Standard Model predictions 
that seem to disagree with experimental measurements. That is the case, for 
example, of $sin(2\beta)$ and $f_{D_s}$. This report discusses 
efforts to resolvethe origin of those discrepancies from the lattice side.}

\FullConference{The XXVI International Symposium on Lattice Field Theory\\
                 July 14-19 2008\\
                 Williamsburg, Virginia, USA}

\begin{document}

\section{Introduction}

\label{introduccion}

The relevance of lattice calculations of heavy quark quantities is derived  
not only from their contribution to the extraction of Standard Model (SM) 
parameters with high precision, but also from their potential to unveil 
New Physics (NP) effects and put constraints on Beyond the 
Standard Model (BSM) theories. They are also a good ground to test 
lattice techniques against well known experimental quantities. 
This program must be carried out together with and having in mind existing 
and future experimental measurements. As an example, the new tagged angular 
analyses of $B_0\to J/\Psi \phi$ by the CDF \cite{CDFBtoJ} and D\O  
\cite{D0BtoJ} collaborations have allowed the extraction of 
the $B^0_s$ mixing phase that can be compared against SM predictions 
that need several inputs from lattice calculations 
\cite{lenznierste06,UTfit08}. The disagreement found in that comparison, 
first pointed out in \cite{lenznierste06}, gives the improvement 
of those lattice inputs primordial importance.

Error analysis is crucial to this effort.  
Experimental errors of most of the relevant quantities are  
now at the few percent level. 
$B^0$ mass differences in both the $B^0_s$ and $B^0_d$ systems are 
known with less than 1\% error \cite{CDFB0mixing,PDG06}. 
The CLEO-c collaboration has recently improved the determination of 
both branching fractions of the $D$ and $D_s$ leptonic decays, including 
the measurement of the $D_s$ decay to $\tau$ \cite{stoneFPCP08}. 
The experimental error in the extraction of the corresponding decay 
constants is now 3-4\%. Semileptonic branching ratios of the $D$ meson 
going to $K$ and $\pi$ allow the extraction $\vert V_{cs}\vert$ and 
$\vert V_{cd}\vert$ with 2\% and 4\% experimental error respectively 
\cite{Skwarnicki}. For the extraction of $\vert V_{ub} \vert$ 
and $\vert V_{cb}\vert$ from $B$ semileptonic decays, experimental 
errors are 6\% and 1.5\%, respectively \cite{babarVub,HFAG08}. 
In the study of heavy flavour observables, the main source of 
uncertainty is thus the error in the theoretical predictions 
of the non-perturbative inputs. In order to be relevant for phenomenology, 
according to the experimental numbers listed above, those 
calculations are needed with accuracy of 5\% or better. 
For this, realistic sea quarks must be included and 
all sources of systematic uncertainty must be addressed, with   
the corresponding errors rigorously estimated. 
In addition, the validity of the Chiral Perturbation Theory (ChPT) 
techniques used for extrapolating to the physical point and 
region of applicability must be checked as discussed at this conference 
by Karl Jansen \cite{jansenLat08} and Laurent Lellouch 
\cite{lellouchLat08}. Now it is possible to perform such calculations 
and a few examples have already appeared as described at this conference 
and listed in this review.

The heavy flavour sector is currently very interesting due to the recent 
claims of possible discrepancies between SM expectations and some 
flavour observables, for example, the disagreement in the value of the 
$B^0_s$ mixing phase mentioned above \cite{HFAGweb08}. 
Two other parameters for which disagreement 
between experiment and SM prediction has been discussed at this 
conference are the decay constant of the $D_s$ meson \cite{BK08} and 
the Unitarity Triangle (UT) angle $\sin(2\beta)$ \cite{LunguiSoni08}. 
All these analyses rely on and are very sensitive to  
lattice calculations of different quantities: decay constants, the 
$SU(3)$ breaking ratio of $B^0$ mixing parameters $\xi$- defined in 
Section \ref{B0mixing}, the form factors needed to extract the 
Cabibbo-Kobayashi-Maskawa (CKM) matrix matrix elements $\vert V_{cb}\vert$ 
and $\vert V_{ub} \vert$ from semileptonic decay experimental data, $\dots$ 
In the following sections I summarize the advances in those calculations, 
including updates and new studies, which will be crucial to understand 
the origin of the discrepancies mentioned above.

\section{Decay constants}

The lattice determination of pseudoscalar decay constants, 
together with experimental measurements of pseudoscalar 
leptonic decay widths, can be used to extract the value 
of the CKM matrix elements involved in the process via 
$\Gamma(P_{ab}\to l \nu) = ({\rm known\,factors})\, f_P^2\vert V_{ab} 
\vert^2$. On the other hand, if the corresponding CKM matrix 
elements are known from other sources, the experimental 
measurements can be used to test lattice QCD calculations.

\subsection{$f_D$ and $f_{D_s}$: The $f_{D_s}$ puzzle} 

\label{fDfDssection}

The charm sector was thought to be able to provide valuable tests 
of lattice QCD techniques, since, in principle, it is not 
expected to be significantly affected by NP or, at least, it is not 
expected to be the first place where NP would show up.  
Important progress has also been made by B-factories and CLEO-c 
to reduce the errors in the 
experimentally measured branching fractions for both $D$ and $D_s$. 
Fixing $V_{cs(d)}=V_{us(d)}$, one can extract the values of 
the decay constants from experiment with an error of around $3-4\%$. 
When describing charm quarks on the lattice, one must take into 
account the fact that it falls in a regime which does not correspond 
to neither the heavy nor the light regimes. Heavy quark effective theories 
are not able to describe charm quantities with the 
high precision needed by phenomenology 
and, on the other hand, relativistic theories need improvements 
in both actions and operators  in order to keep cut-off effects under 
control and get those needed precisions.

The FNAL/MILC collaboration has presented at this conference the 
status of the reanalysis of their existing data for the decay constants 
in the charm and the bottom sectors, and their plans for future 
runnings \cite{fDfnal}. The light quarks in this analysis are simulated 
with improved staggered quarks, in particular, the Asqtad 
action \cite{asqtad}, and the heavy quarks are simulated following 
the Fermilab approach \cite{kkm}. The simulations are performed on  
MILC configurations with $N_f=2+1$ flavours of sea quarks. Three lattice 
spacings have been analyzed, $a=0.15,0.12,0.09~fm$, with three or 
five sea quark masses at each lattice spacing 
and between nine and twelve light 
valence quark masses at each sea quark mass choice. The smallest  
light sea quark mass simulated is $m_l\simeq m_s/10$. 

An important feature of this analysis 
is the use of a partially non-perturbative method to renormalize the 
currents. This reduces the uncertainty in the renormalization procedure 
compared to the one-loop perturbative approach by a factor of between 
2 and 3 (depending on the lattice spacing). 
Another important feature is the use of Staggered Chiral Perturbation 
Theory (SChPT) \cite{schpt} to simultaneously extrapolate 
the results to the continuum 
and the physical masses. SChPT allows to remove the dominant 
light discretization effects, since the expressions explicitly account for 
the dominant taste-changing violating effects and dependence 
on the lattice spacing. The downside is that the SChPT expressions are 
rather complicated and depend on a large number of parameters that must 
be fixed using other simulations or as an output of the extrapolation 
fits. These two approaches are common to most of 
all FNAL/MILC calculations mentioned 
in this paper except for the $B^0-\bar B^0$ study for which the 
applicability of the partially non-perturbative renormalization is still 
under investigation. In those analyses at least NLO SChPT expressions are 
considered and, in most of the cases, analytic NNLO terms are also 
included in the fits.

\begin{figure}
\begin{center}
{\includegraphics[width=0.43\textwidth,]{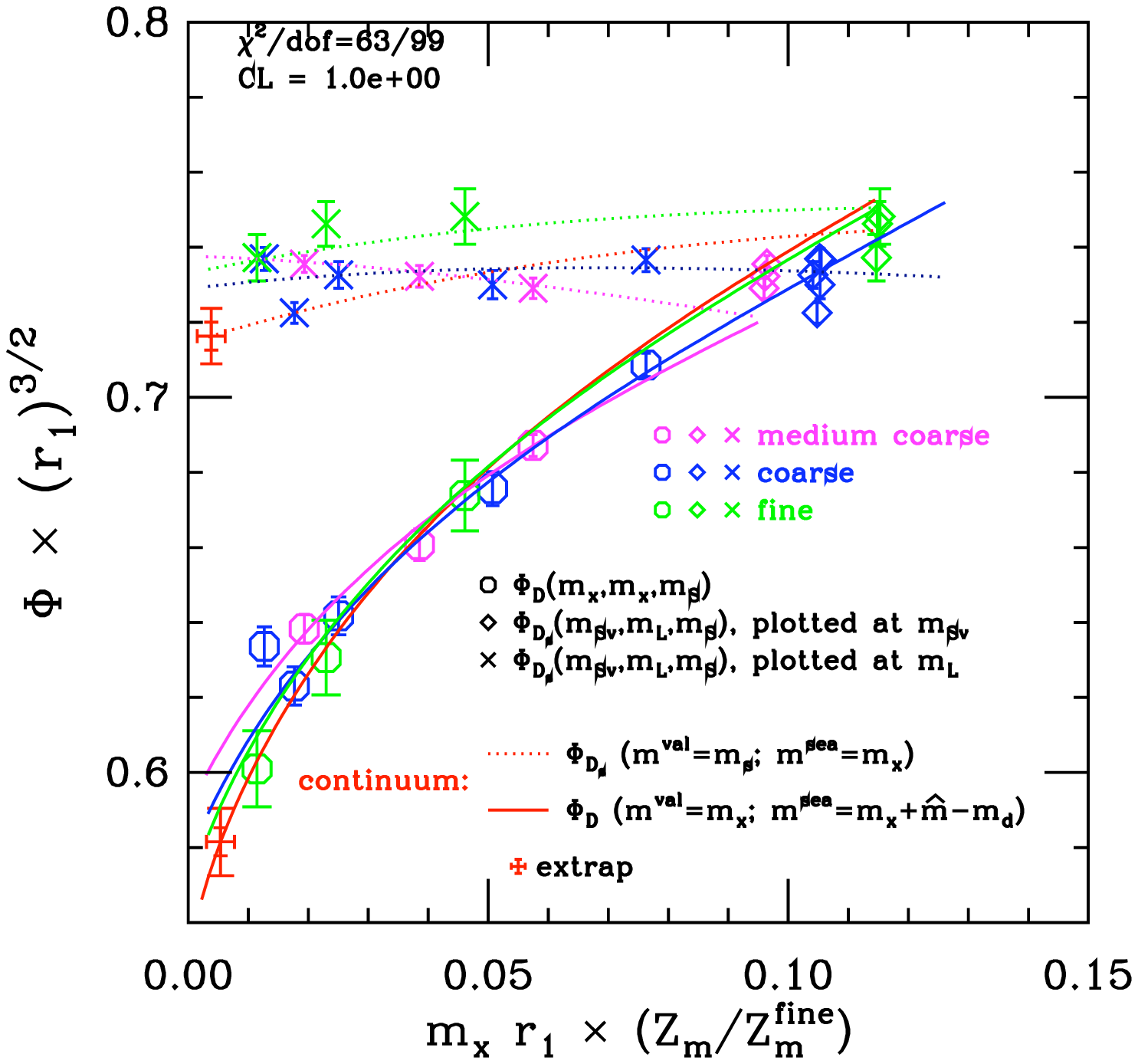}}
{\includegraphics[width=0.43\textwidth]{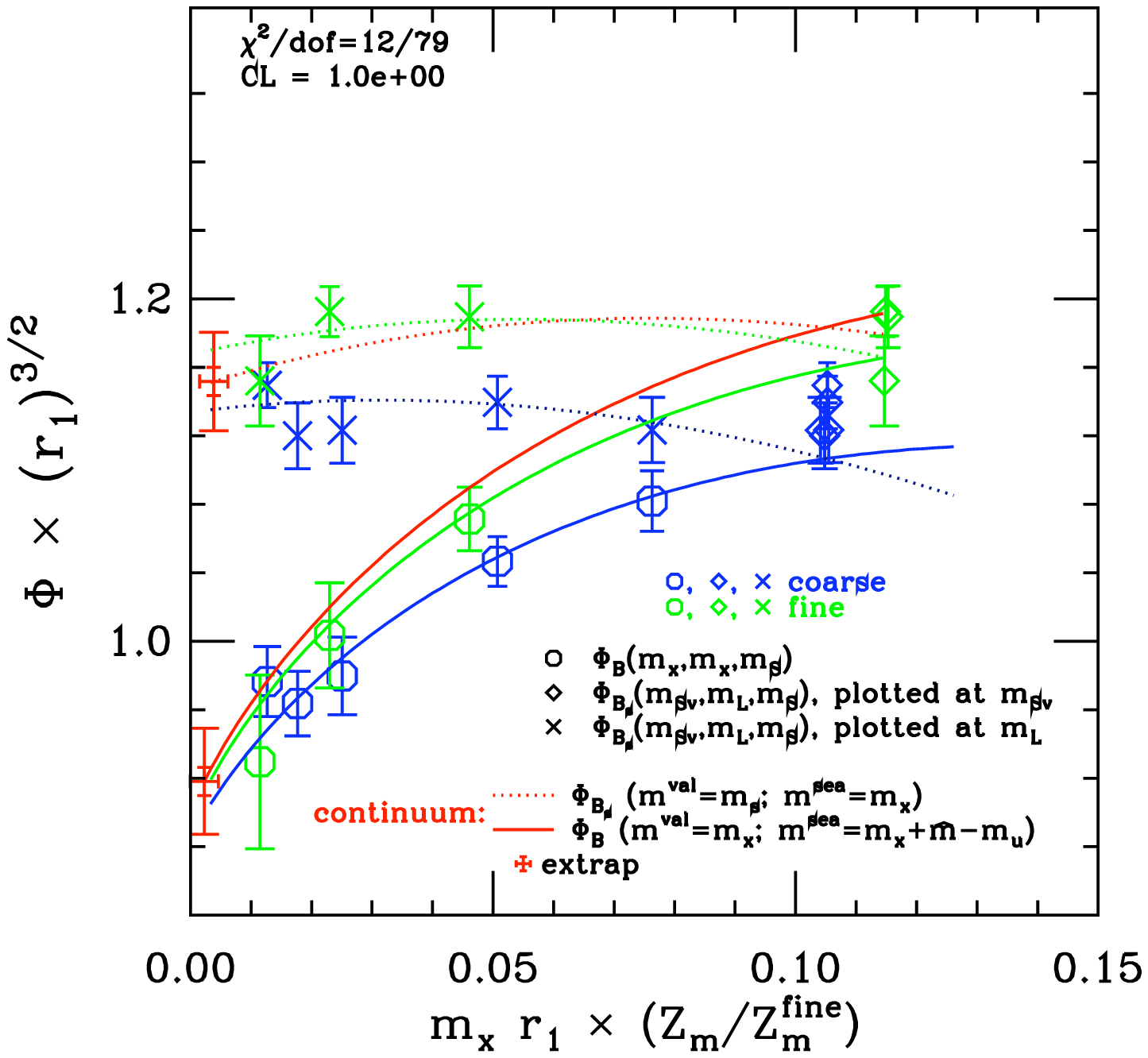}}
\caption{$\Phi=f_{D_q}\sqrt{M_{D_q}}$ and $\Phi=f_{B_q}\sqrt{M_{B_q}}$ 
in $r_1$ units. Hexagons correspond to the $D/B$ parameters plotted 
as a function of the light valence quark mass with the sea masses 
fixed to the physical values. Crosses 
correspond to $D_s/B_s$ parameters plotted as a function of 
the sea light quark mass.  Figure taken from \cite{fDfnal}.  
\label{fig:fDfnal}}
\end{center}
\end{figure}

The extrapolated values of $f_{D}$ and $f_{D_s}$ are extracted from the 
same fit, whose results are shown in Figure \ref{fig:fDfnal}. 
The dependence on the light sea quark mass is very mild as 
can be seen in that figure. 
The light quark masses are light enough so the lattice 
data show the correct behaviour dictated by the logarithms in the 
chiral expressions. The results obtained for the two decay constants and 
the ratio are
\ba \label{fDfnalresults}
f_D= 207(11){\rm MeV}\quad f_{D_s}=249(11){\rm MeV} \quad
f_{D_s}/f_D=1.200(27)\,.
\ea
The errors in (\ref{fDfnalresults}) are dominated by the tuning of 
$m_c$ and discretization errors. They are planning to improve 
this calculation by performing simulations at $a=0.06~fm$, 
as  well as on the extra configurations generated 
on the existing ensembles. In addition, they are currently 
retunning the charm quark mass. These improvements can be 
applied to all the FNAL/MILC analyses discussed in this paper.

The FNAL/MILC results can be compared with the HPQCD calculation in 
\cite{fDhpqcd} that employs the same $N_f=2+1$ MILC gauge configurations. 
The main difference between the two calculations is the treatment 
of the valence quarks, described by the HISQ action \cite{hisq} in 
the HPQCD analysis. Other differences are the fact that HPQCD has  
no need of renormalization (they use PCAC relations to extract 
the decay constants) and instead of using SChPT to perform 
the extrapolation to the continuum 
and the physical masses they perform a Bayesian analysis including 
continuum NLO ChPT plus different $\order(a^2)$ functional forms, and 
second and third order polynomial in the masses. In addition, the 
HPQCD collaboration only studied full QCD points. Their results 
\ba
f_D=(208\pm4){\rm MeV} \quad f_{D_s}=(241\pm 3){\rm MeV} \quad 
f_{D_s}/f_D=1.162(9)\,
\ea
agree very well with the FNAL/MILC ones, although with smaller errors 
due mainly to the fact that the valence quark action for the $b$ quark 
is more improved. 

\begin{figure}
\vspace*{-0.5cm}
\begin{center}
\includegraphics[width=0.35\textwidth,angle=-90]{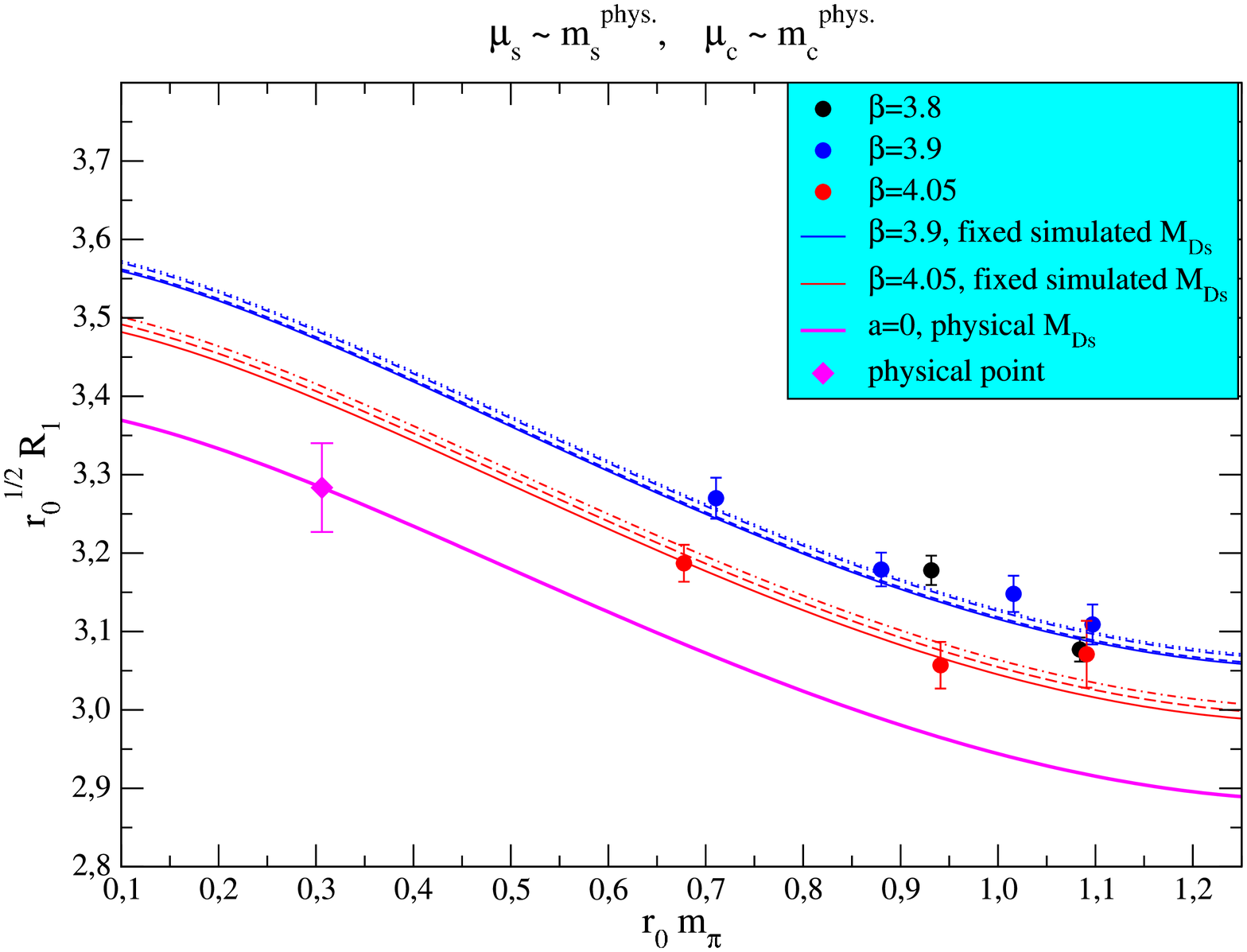}
\end{center}
\caption{Chiral extrapolation for the ratio $R_1$ defined in 
the text as a function of $r_0 m_\pi$ for the three lattice 
spacings analyzed and the corresponding continuum extrapolation. Figure 
is taken from \cite{TarantinoLat08}. \label{fig:fdetmc}}
\end{figure}

Another interesting on-going calculation, albeit less complete, 
of decay constants in the 
charm sector is the one by the ETM Collaboration whose preliminary 
results have been presented at this conference \cite{TarantinoLat08}. 
They employ a twisted mass (tmQCD) formalism with $N_f=2$ light 
sea quarks. The hadronic quantities like meson masses and decay constants 
are automatically $\order(a)$ improved using this formalism at maximal 
twist \cite{tmimprov}. In addition, quark masses are 
multiplicatively renormalized (and the renormalization factor can be 
determined non-perturbatively) and decay constants do not require to be 
renormalized. They are extracted from PCAC relations. In this study 
full QCD points for light quark masses between $m_s/5$ and $m_s/2$  
are analyzed for three lattice spacings, $a=0.1,0.0855,0.0667~fm$. Simulations 
are performed for several values of the charm quark mass around the 
physical one. 
A good way of doing the chiral extrapolation milder is by considering 
ratios of decay constants. The ETM collaboration use the ratios 
$R_1=f_{D_s}/f_K$ and $R_2=\left[f_{D_s}\sqrt{M_{D_s}}/f_K\right]/
\left[f_{D}\sqrt{M_{D}}/f_\pi\right]$, and their own results for 
$f_K$ and $f_K/f_\pi$ to extract the decay constants in the charm sector. 
The chiral extrapolation is performed together with the extrapolation 
to the continuum through a simultaneous fit analogous to the ones 
employed by the HPQCD collaboration and described above. 
Together with the continuum 
NLO Heavy Meson ChPT expressions (HMChPT)\footnote{The authors in 
\cite{TarantinoLat08} 
claim that the strange quark mass could be too heavy to be described 
by ChPT, so they use SU(2) HMChPT expressions (obtained from the 
SU(3) ones by performing and expansion in $m_l/m_s$) and use 
SU(3) expressions to estimate the corresponding systematic error.} 
explicit dependence in $a^2$ is also included in the functional form.  
The results from this simultaneous fit for $R_1$ are shown in Figure 
\ref{fig:fdetmc}. 
The preliminary results presented in \cite{TarantinoLat08} are
\ba \label{fdresultsetmc}
f_{D}=(197\pm 7\pm 12){\rm MeV} \quad f_{D_s}=(244\pm 4\pm 11){\rm MeV} \quad 
f_{D_s}/f_D=(1.24\pm0.04\pm 0.02)\,,
\ea
where the first error is statistical and the second systematic. As can be 
seen in Figure \ref{fig:fdetmc}, 
there is a significant gap between the results for 
the smallest lattice spacing and the continuum extrapolated numbers, which 
translates into a sizable uncertainty associated with the 
continuum extrapolation. This analysis needs a better control over 
discretization errors. The 
collaboration is planning to perform simulations at a smaller lattice 
spacing $a\simeq 0.06~fm$ to address that limitation. The results in 
(\ref{fdresultsetmc}) agree with the ones quoted by the collaborations  
simulating $N_f=2+1$ sea quarks. However, the effects of including a 
third sea quarks are missing and the systematic analysis does not include 
the corresponding error.

\begin{figure}[th]
\begin{center}
\includegraphics[width=0.30\textwidth,angle=-90]{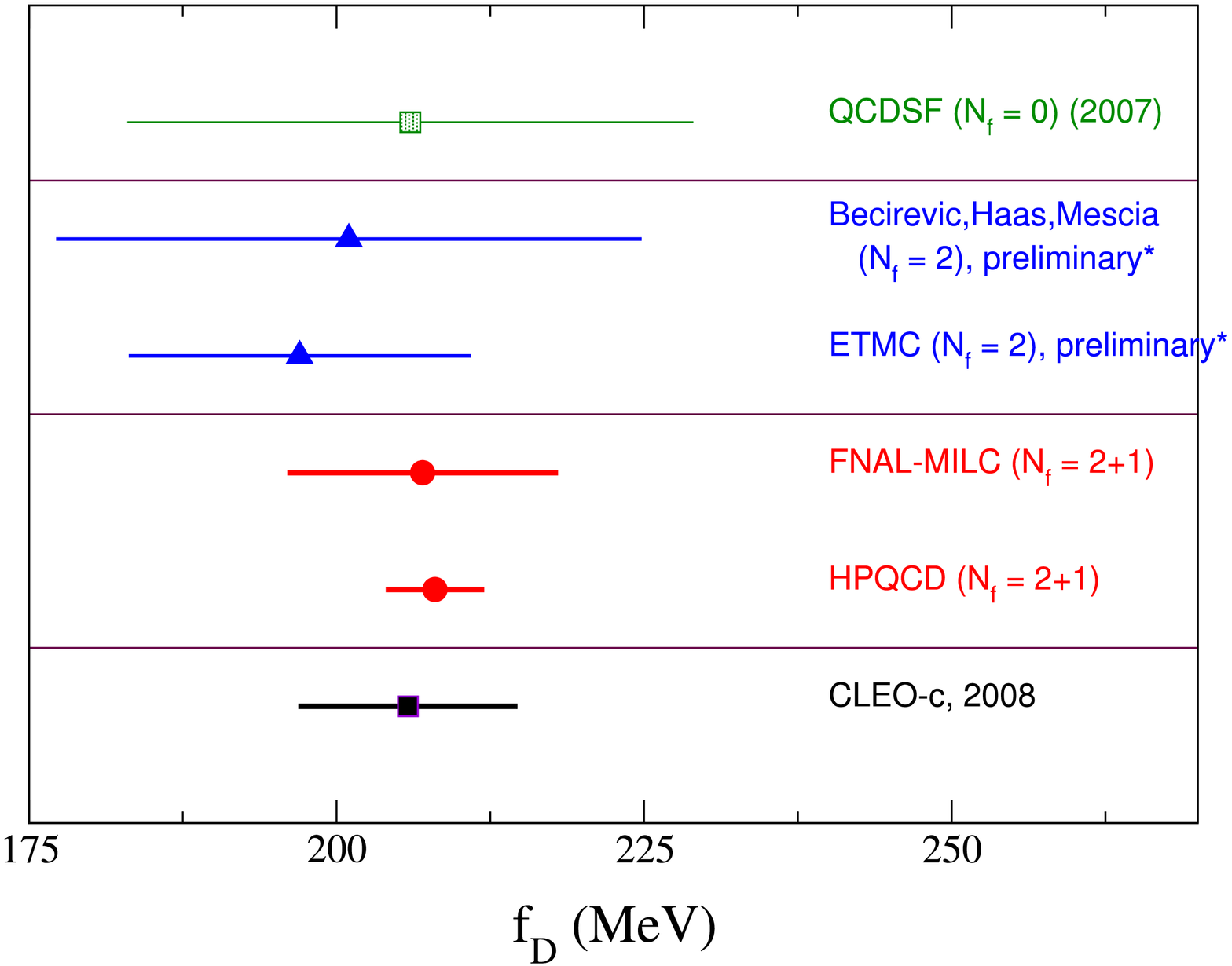}
\hspace*{1.2cm}\includegraphics
[width=0.30\textwidth,angle=-90]{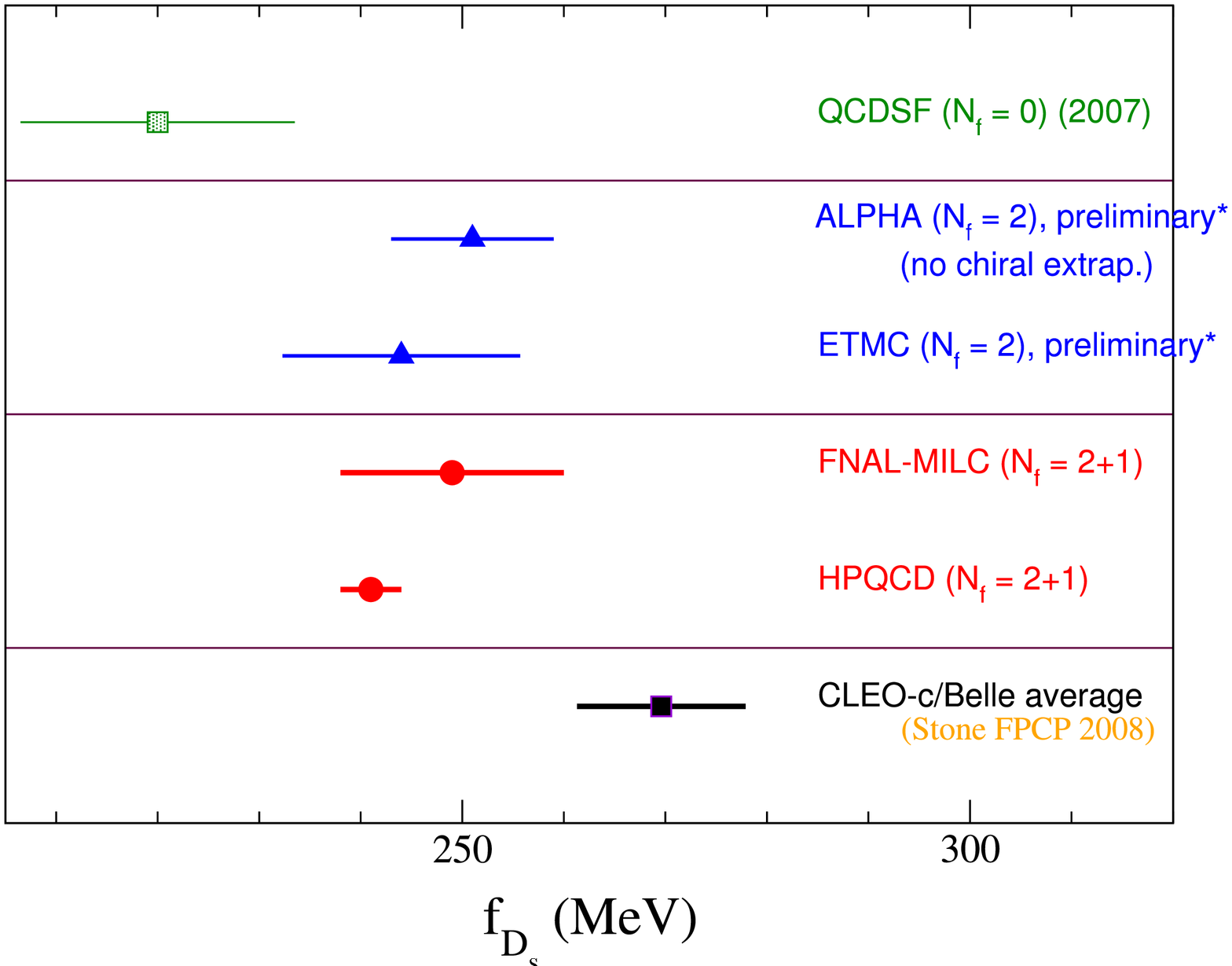}
\end{center}
\caption{Values of $f_{D}$ and $f_{D_s}$ from experiment 
\cite{cleoc08,Bellefds} and recent lattice calculations. \label{fdandfds}}
\end{figure}

If we take the most recent lattice results for $f_D$ and $f_{D_s}$ and 
compare then with experiment -as has been done in Figure 
\ref{fdandfds}-, it can be seen that the agreement for $f_D$ 
is very good, but for $f_{D_s}$ there is a clear tendency for lattice 
calculations to give smaller values than experiment. In particular, 
the HPQCD result, which is the most accurate one, is more than 
$3\sigma$ away from experiment. 
The fact that the rest of quantities calculated by the HPQCD calculation 
with the same actions, configurations, input parameters, etc ($f_D$, 
$f_K$, $f_\pi$, $m_D$, $m_{D_s}$, $\frac{2m_{D_s}-m_{\eta_c}}
{2m_D-m_{\eta_c}}$, $\dots$) agree with experiment at the 2\% level, 
strongly supports the reliability of their results for $f_{D_s}$. 
In any case, it would be good to have an independent $N_f=2+1$ 
lattice calculation with comparable errors of this quantity. 
On the experimental side, it would be also interesting that some issues 
like the use of unitarity to fix the value of $V_{cs}$ or 
the more careful treatment of radiative corrections were addressed. 
On the other hand, it is possible that this discrepancy is 
indicating the presence of NP effects as discussed at  
this conference \cite{BK08}.

The ALPHA collaboration has also presented preliminary results for $f_{D_s}$ 
as well as the charm quark mass at this conference \cite{vonHippelLat08}. 
Those are simulations with $N_f=2$ Wilson fermions which seems to indicate 
that cutoff effects are under control due to the small lattice spacings used 
(finest lattice spacing is $a=0.04~fm$).

\subsection{$f_B$ and $f_{B_s}$}

\label{sec:fbs}

The decay constants for leptonic $B$ and $B_s$ mesons are also 
parameters of phenomenological relevance. Lattice results for the 
decay constants in the $B-$meson sector are needed more than in the 
$D-$meson sector since the corresponding CKM matrix elements.
The $B$ leptonic decays themselves would also be a sensitive 
probe of effects from charged Higgs bosons. The $B$ decay constants 
are also used in the SM predictions for processes very 
sensitive to beyond SM effects, such as $B_s\to\mu^+\mu^-$.

One of the advantages of the Fermilab action is that it can be 
efficiently used to describe both charm and bottom quarks. 
This has allowed the FNAL/MILC collaboration to use the same setup as the 
one described in section (\ref{fDfDssection}) to calculate $f_B$, 
$f_{B_s}$, and the ratio of both decay constants. 
The results from the simultaneous chiral and continuum extrapolations for  
$f_B$ and $f_{B_s}$ are shown in Figure \ref{fig:fDfnal}.  
After the extrapolations the numbers obtained are 
$f_B = (195\pm 11){\rm MeV}$, $f_{B_s}=(243\pm 11){\rm MeV}$ and  
$f_{B_s}/f_B = 1.25\pm 0.04$. Statistics is in this case a more 
important source of error than for $f_D,f_{D_s}$, while in the 
$B$ sector results are less sensitive to the tuning of the $b$ quark mass.

These results agree with the $N_f=2+1$ calculation of the HPQCD 
collaboration \cite{fBHPQCD} using also the MILC configurations 
but with $b$ quarks simulated with the NRQCD action in \cite{nrqcd}: 
$f_B = (216\pm 22){\rm MeV}$, $f_{B_s}=(260\pm 26){\rm MeV}$ and 
$f_{B_s}/f_B = 1.20\pm 0.03$. The errors in HPQCD numbers are larger 
by around a factor of two mainly due to the uncertainty associated with the 
one-loop perturbative renormalization used for the matching of 
the corresponding currents.

There have been two recent quenched calculations of $f_{B_s}$ using 
$\order(a)$ improved Wilson fermions but following different approaches, 
both going beyond the static approximation. The authors in \cite{fBsALPHA08} 
performed simulations using a relativistic description with heavy masses around 
the charm quark mass together with simulations in the static approximation. 
They interpolated the results to the physical point using the expression 
$r_0^{3/2}\frac{F_{PS}\sqrt{m_{PS}}}{C_{PS}(M/\Lambda_{\overline{MS}})}
=A\left(1+\frac{B}{r_0m_{PS}}\right)$. The slope leads to a $1/m^2_{PS}$  
correction of about 10\% at the physical point. The number obtained 
is $f_{B_s}=191(6)$, where all errors except quenching are considered. 
The other quenched analysis, described in 
\cite{GST08}, is based on the use of the Step Scaling Method (SSM) \cite{ssm} 
and the Heavy Quark Effective Theory (HQET). The authors calculate the 
SS functions for several heavy masses around the charm quark mass and in the 
static approximation. They then extract the  final by performing an 
interpolation in $1/m$ to the physical $B^0_s$ mass. As expected, SS 
functions were found to be very weakly dependent on the mass. 
The result obtained 
is $f_{B_s}=193(7)$. In this analysis, as well as the one 
in \cite{fBsALPHA08}, the inclusion of the static point improves 
noticeably the control over the heavy quark mass dependence. 
Both numbers in \cite{fBsALPHA08} and \cite{GST08} are in very good agreement, 
which gives confidence to the systematic error analyses. It is clear from 
the results obtained in these two calculations and those from the $N_f=2+1$ 
calculations mentioned above that the effect of the 
quenched approximation is quite noticeable in $f_{B_s}$. However,  
the quenched results already show that the 
two methods are promising to get accurate values of $f_{B_s}$.
It would be also interesting to see whether these two results agree with 
a pure HQET calculation including $1/M$ corrections.

Several collaborations working on static-light studies have presented 
preliminary results at this conference. The RBC/UKQCD collaboration has 
completed the one-loop renormalization calculations needed for its 
$f_B$ and $B^0-\bar B^0$ analyses with $N_f=2+1$ sea quarks 
and domain wall light quarks \cite{tomomiLat08}. The ETMC collaboration 
is using twisted mass fermions with $N_f=2$ together with static 
bottom quarks. They have presented first results for the spectrum of 
$B_s-$mesons \cite{WagnerLat08}. And beyond the static limit, the 
HPQCD collaboration reported also at this conference preliminary results 
for the calculation of $B_c$ and $B_s$ meson masses with NRQCD and HISQ 
fermions \cite{GregoryLat08}.

\section{Semileptonic decays}

Semileptonic decays of heavy-light mesons are currently used to 
extract the CKM matrix elements $\vert V_{cb}\vert$, $\vert V_{ub}\vert$, 
$\vert V_{cd}\vert$ and $\vert V_{cs}\vert$. The theory input needed to 
get those parameters from experimentally measured semileptonic widths are 
the form factors in terms of which the hadronic matrix elements 
involved on those decays are parametrized. For example, for the decay   
$D\to K l\nu$, the differential decay rate is given by 
$\frac{d\Gamma}{dq^2} = ({\rm known\, factors})\vert V_{cs}
\vert^2 f_+^2(q^2)$, where $f_+(q^2)$ is the vector form factor. 
This vector form factor can be extracted from the 
matrix element of the vector current 
$\langle K\vert V^\mu\vert D\rangle  = f_+(q^2)(p_D+p_K-\Delta)^\mu
+f_0(q^2)\Delta^\mu$ , with $\Delta^\mu=(m_D^2-m_K^2)q^\mu/q^2$ 
and $q=p_D-p_K$. Analogously, the processes 
with a vector meson in the final state are described by four form factors.

There are several lattice techniques that are considerably contributing,  
or will do in the near future, to the improvement of the calculations
of semileptonic form factors. An important reduction 
of the errors is got by using double ratios methods \cite{doubleratios}, 
whose goal is the cancellation of statistical and systematic errors 
as well as chiral corrections between denominator and numerator. 
The use of ratios often also yields a milder chiral extrapolation.  
Another important ingredient to be considered when 
calculating form factors is the parametrization used to describe the 
$q^2$ dependence. Traditional lattice QCD methods are able to 
calculate form factors in an accurate way only at 
high values of $q^2$, while experimental values are more precise for 
low $q^2$. One thus needs a description of the $q^2$ dependence 
to connect both set of results. The optimal choice is a model 
independent parametrization, as discussed later in section \ref{sec:Btopi}. 
On the other hand, with twisted boundary conditions \cite{twboundary}   
the simulations can be performed at smaller recoil momenta than 
with periodic boundary conditions, which allows a better resolution 
of the small recoil region. 

Another approach to deal with the large discretization errors at small 
values of $q^2$ is using a fermion formulation that address directly 
the issue. This is the case of moving nonrelativistic QCD (mNRQCD), with  
which discretization errors can be kept under control for small values of 
$q^2$. Preliminary results for the first calculation of form factors using 
mNRQCD have been presented at this conference \cite{MeinelLa08}. The 
statistical errors are still large, but the authors will improve them by 
using random wall sources.

\subsection{ Exclusive $B\to D^*(D) l \nu$:   
determination of $\vert V_{cb}\vert$}

The reduction in the error of $\hat B_K$ under 10\% 
\cite{lellouchLat08} has made the error in the determination of the 
CKM matrix element $\vert V_{cb}\vert$ to become a dominant source 
of uncertainty in the analysis of $\varepsilon_K$, which measures indirect 
CP violation in the Kaon system. This CKM matrix element is also needed 
in the analysis of some rare kaon decays, for example, 
$K\to \pi\nu\bar\nu$.

There are two processes that have been used to extract the value of 
$\vert V_{cb}\vert$ using lattice techniques,  
$B\to D l \nu$ and  $B\to D^* l \nu$. In general, 
the analysis of these processes depend on a combination of 
four form factors which are functions of $\omega$ (the scalar product 
of the velocities of the meson in the initial and the final state), but 
at zero recoil ($\omega=1$) only one form factor, $h_{A_1}$, is 
needed. The experimental results for $B\to D^* l \nu$ at zero recoil 
have smaller errors than those for $B\to D l \nu$, so it yields 
a more precise determination of $\vert V_{cb}\vert$.

\begin{figure}[th]
\vspace*{0.2cm}
\begin{center}
{\includegraphics[width=0.42\textwidth]{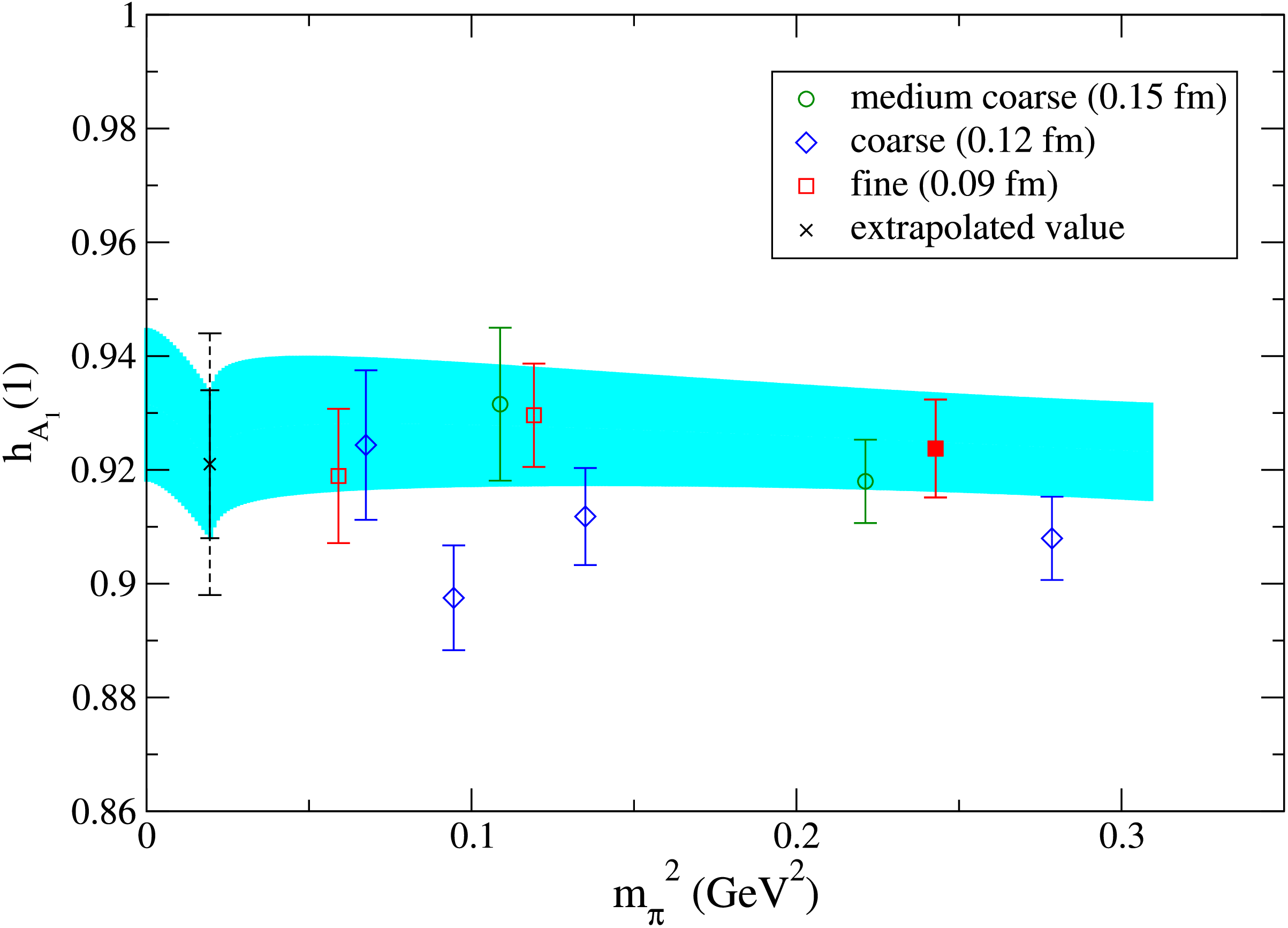}}
\hspace*{1.2cm}
{\includegraphics[width=0.42\textwidth]{./Dibujos/B2Dstar_a2}}
\end{center}
\caption{The figure in the left shows the full QCD points for the form 
factor $h_{A_1}(1)$ versus $m_\pi^2$ on the three lattice spacings. The curve 
is already extrapolated to the physical strange sea quark mass. The cross is 
the extrapolated value with statistical (solid line) and systematic 
(dashed line) errors. The figure in the right shows $h_{A_1}(1)$ 
at physical quark masses versus $a^2(fm^2)$. Figures taken 
from \cite{Laiho08}. \label{fig:BtoD*}}
\end{figure}

The FNAL/MILC collaboration presented last year at this conference 
\cite{LaihoLat07} a preliminary result for its calculation of $h_{A_1}$.   
This analysis has recently been completed \cite{Laiho08}. 
They perform a $N_f=2+1$ calculation 
with Asqtad staggered light quarks and heavy quarks described with 
the Fermilab formalism. The analysis introduces a new double ratio 
method which gives the form factor at zero recoil directly. There is 
also a reduction of the computational cost from previous FNAL/MILC analyses 
\cite{oldBtoD*} since the new method does not require simulations 
at several heavy quark masses. The relation between the double ratio and the 
form factor, 
\be
\vert {\cal F}_{B\to D^*}(1) \vert^2 = \frac{\langle D^*\vert \bar c 
\gamma_j\gamma_5 b\vert\bar B\rangle \langle \bar B\vert \bar b 
\gamma_j\gamma_5 c\vert D^*\rangle }
{\langle D^*\vert \bar c \gamma_4 c\vert D^*\rangle 
\langle \bar B\vert \bar b \gamma_4 b\vert\bar B\rangle }
\, ,
\ee
is  exact to all orders in the heavy-quark 
expansion in the continuum. Statistical errors in the numerator and 
denominator are highly correlated and largely cancel. Most 
of the renormalization also cancels, yielding a small uncertainty for the 
perturbative matching. Figure \ref{fig:BtoD*} shows the mass and lattice 
spacing dependence of the data. In both cases the extrapolation needed 
to go to the physical/continuum limit is very mild. 
The final result obtained for the form factor after chiral and 
continuum extrapolation is $h_{A_1}(1)=0.921
\pm0.013\pm0.021$ \cite{Laiho08}, where the first error is 
statistical and the second one includes all sources of systematic 
errors and it is dominated by heavy-quark discretization errors. The 
CKM matrix element $\vert V_{cb}\vert$ extracted from 
this value of the form factor and the experimental averages in 
\cite{HFAG08} is $\vert V_{cb}\vert = (38.8 \pm 0.6_{exp} \pm 1.0_{theo})
\times 10^{-3}$. This value differs by $2\sigma$ from the one extracted from 
inclusive decays \cite{PDG06},  $\vert V_{cb}\vert = (41.7 \pm 0.7)
\times 10^{-3}$. In this FNAL/MILC calculation all the sources of error are 
under control, none of them being larger than 1.5\% \cite{Laiho08}, 
giving to a very clean extraction of  $\vert V_{cb}\vert$.

\begin{figure}
\begin{center}
\includegraphics[width=0.49\textwidth]{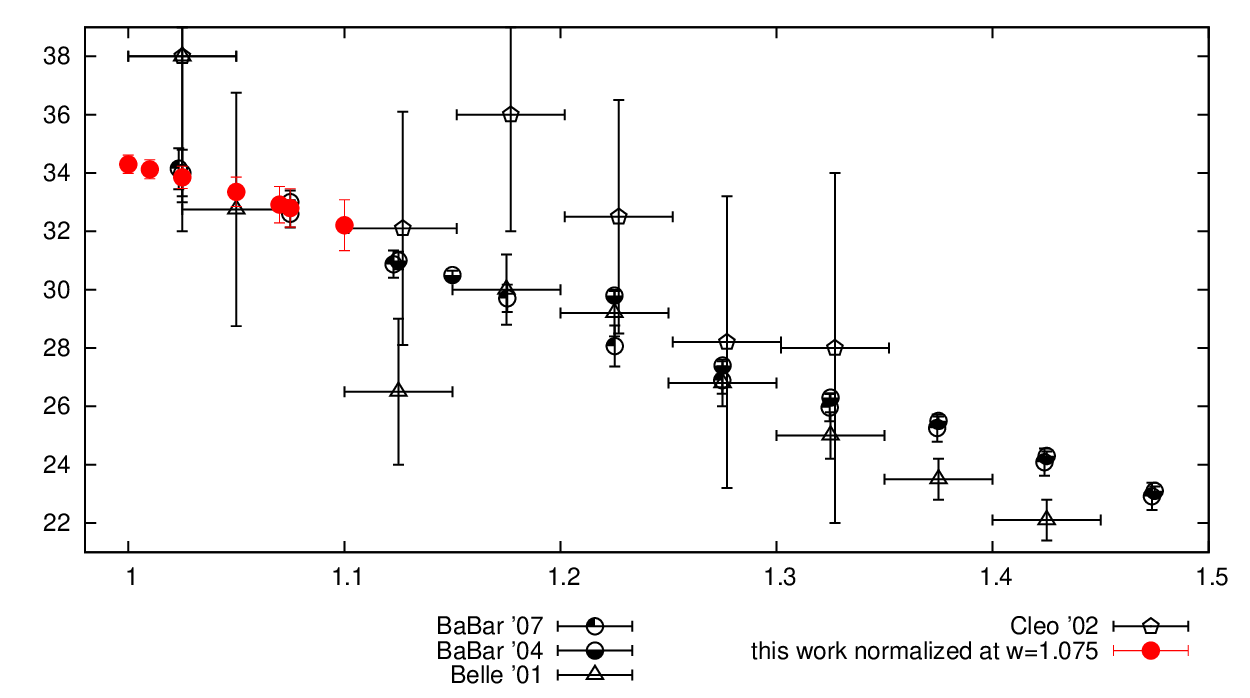}
\includegraphics[width=0.49\textwidth]{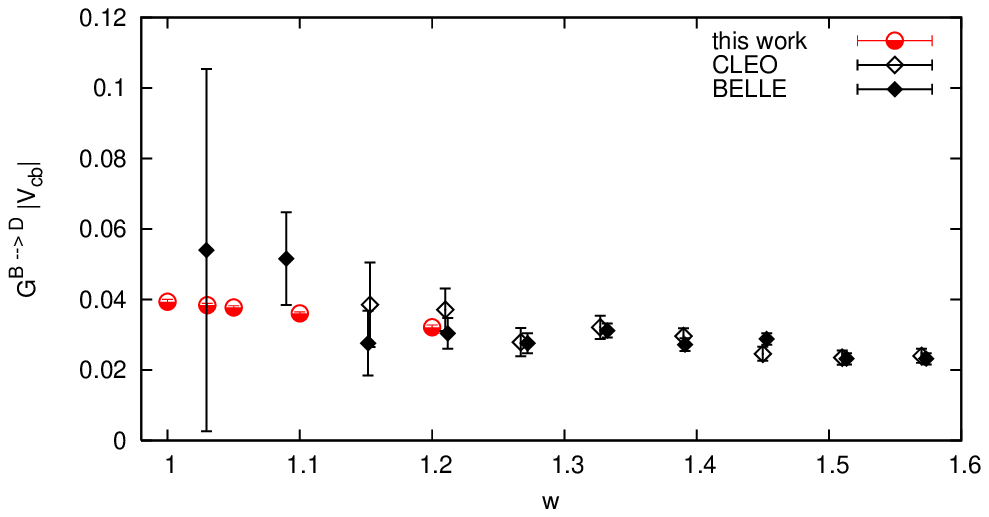}
\end{center}
\caption{Comparison of experimental data and 
quenched lattice data for $B\to D^*$ form factors \cite{Divitiis08} 
and $B\to D$ form factor \cite{Divitiis07} respectively. The value of 
$\vert V_{cb}\vert$ in this plot lattice is extracted from matching 
experimental and lattice data at $\omega=1.075$ for $B\to D^*$ and 
$\omega=1.2$ for $B\to D$. Figures from \cite{Divitiis08} 
and \cite{Divitiis07}.
\label{BtoDDivitiis}}
\end{figure}

The recent quenched study in \cite{Divitiis08} calculates the combination of 
form factors needed to describe $B\to D^*l\nu$ for values of $\omega\ge 1$, 
$F^{B\to D^*}(\omega)$, using the Step Scaling method. These values 
of $\omega$ can be reached thanks to the use of  twisted flavour 
boundary conditions. Although it is a 
quenched calculation, it is useful to check the rest of the systematics 
entering in the analysis and to discuss some technical issues, 
having in mind the future inclusion of vacuum polarization effects. 
The results obtained for the product 
$F^{B\to D^*}\vert V_{cb}\vert$ together with different experimental 
measurements are shown in the left hand side of Figure \ref{BtoDDivitiis}.
There is a very good overlap between lattice and experimental data 
for the region of $\omega$ studied, but in this process 
it is not so important  
to go to larger values of $\omega$ since for $\omega=1$ experimental 
data are already good. That is however crucial for the decay   
$B\to D l\nu$ where, as can be seen in the right hand side of Figure 
\ref{BtoDDivitiis},  experimental data at zero recoil are very noisy. 
That figure corresponds to the quenched analysis of $B\to Dl\nu$ 
in \cite{Divitiis07}, which follows  the same methodology 
than the one for $B\to D^* l\nu$ in \cite{Divitiis08}. An 
interesting feature of the analysis in \cite{Divitiis07} is that the authors 
determine the ratio $\Delta^{D\to B}(\omega)$, which parametrizes the 
difference between $B\to D e(\mu)\nu_e(\nu_\mu)$ and $B\to D \tau\nu_\tau$. 
This ratio could be extracted from the experimental measurement of 
$d\Gamma\left(B\to D \tau\nu_\tau\right)/d\Gamma\left(B\to D e(\mu)
\nu_e(\nu_\mu)\right)$ and it is independent of CKM inputs, so it potentially 
constitutes a good way to check lattice techniques. The calculation of 
 $\Delta^{D\to B}(\omega)$ is also relevant because the ratio of partially 
integrated rates $Br\left(B\to D \tau\nu_\tau\right)/
Br\left(B\to D e\nu_e\right)$ has been claimed to be a good place to look 
for charged Higgs contributions to low energy observables \cite{KM08}. 
Having results for both channels makes also possible to perform lepton-flavour 
universality checks on the extraction of $\vert V_{cb}\vert$.

\subsection{ Exclusive $B \to \pi l \nu$: 
determination of $\vert V_{ub}\vert$}
\label{sec:Btopi}

The decay $B\to \pi l\nu$ provides a determination of $\vert V_{ub}\vert$ 
that is competitive with inclusive $b\to u$ decays. The main problem 
in studying the exclusive channel with lattice techniques is the 
poor overlap in the momentum transfer, $q^2$, between experimental 
and lattice data, which inflates the final error. A solution to  
this problem is using a model independent parametrization of the shape 
of the form factor. This allows the direct comparison of experimental 
and lattice data even with a poor $q^2$ overlap. 

\begin{figure}
\begin{center}
\begin{minipage}[c]{0.48\textwidth}
	\psfrag{title}[b][b][0.9]{}
	\psfrag{subtitle}[b][b][0.9]{$\chi^2/\text{d.o.f.}=0.32$}
	\psfrag{x-axis}[t][t][0.9]{\begin{scriptsize}$(r_1 E_\pi)^2$\end{scriptsize}}
	\psfrag{y-axis}[b][b][0.9]{\begin{scriptsize}$r_1^{-1/2} f_\perp$\end{scriptsize}}
	\psfrag{0062}[l][l][0.9]{\begin{scriptsize}$a m_l / a m_s = 0.0062/0.031$ fine\end{scriptsize}}
	\psfrag{0124}[l][l][0.9]{\begin{scriptsize}$a m_l / a m_s = 0.0124/0.031$ fine\end{scriptsize}}
	\psfrag{005}[l][l][0.9]{\begin{scriptsize}$a m_l / a m_s = 0.005/0.05$ coarse\end{scriptsize}}
	\psfrag{007}[l][l][0.9]{\begin{scriptsize}$a m_l / a m_s = 0.007/0.05$ coarse\end{scriptsize}}
	\psfrag{01}[l][l][0.9]{\begin{scriptsize}$a m_l / a m_s = 0.01/0.05$ coarse\end{scriptsize}}
	\psfrag{02}[l][l][0.9]{\begin{scriptsize}$a m_l / a m_s = 0.02/0.05$ coarse\end{scriptsize}}
	\psfrag{fit}[l][l][0.9]{\begin{scriptsize}continuum QCD\end{scriptsize}}
	\hspace*{-0.5cm}\rotatebox{270}{\includegraphics[width=0.875\textwidth]{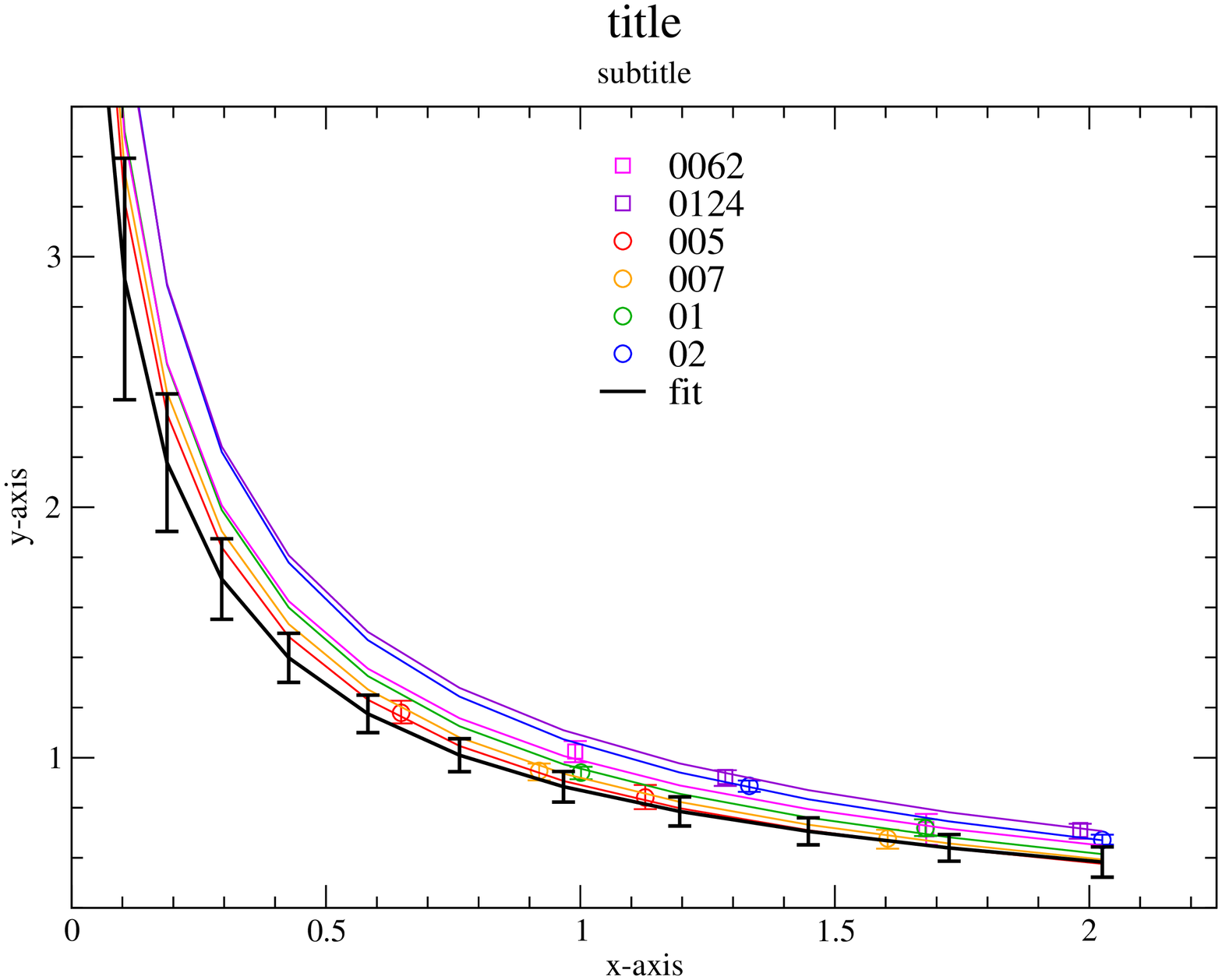}}
\end{minipage}
\begin{minipage}[c]{0.48\textwidth}
	\psfrag{title}[b][b][0.9]{$\chi^2/\text{d.o.f.} = 0.59$}
	\psfrag{x-axis}[t][t][0.9]{$z$}
	\psfrag{y-axis}[b][b][0.9]{$P_+ \phi_+ f_+$}
	\psfrag{set 1}[l][l][0.8]{simultaneous 4-parameter $z$-fit}
	\psfrag{set 2}[l][l][0.8]{Fermilab-MILC lattice data}
	\psfrag{set 3}[l][l][0.8]{BABAR data rescaled by $|V_{ub}|$ from
$z$-fit}
	\rotatebox{270}{\includegraphics[width=0.875\textwidth]{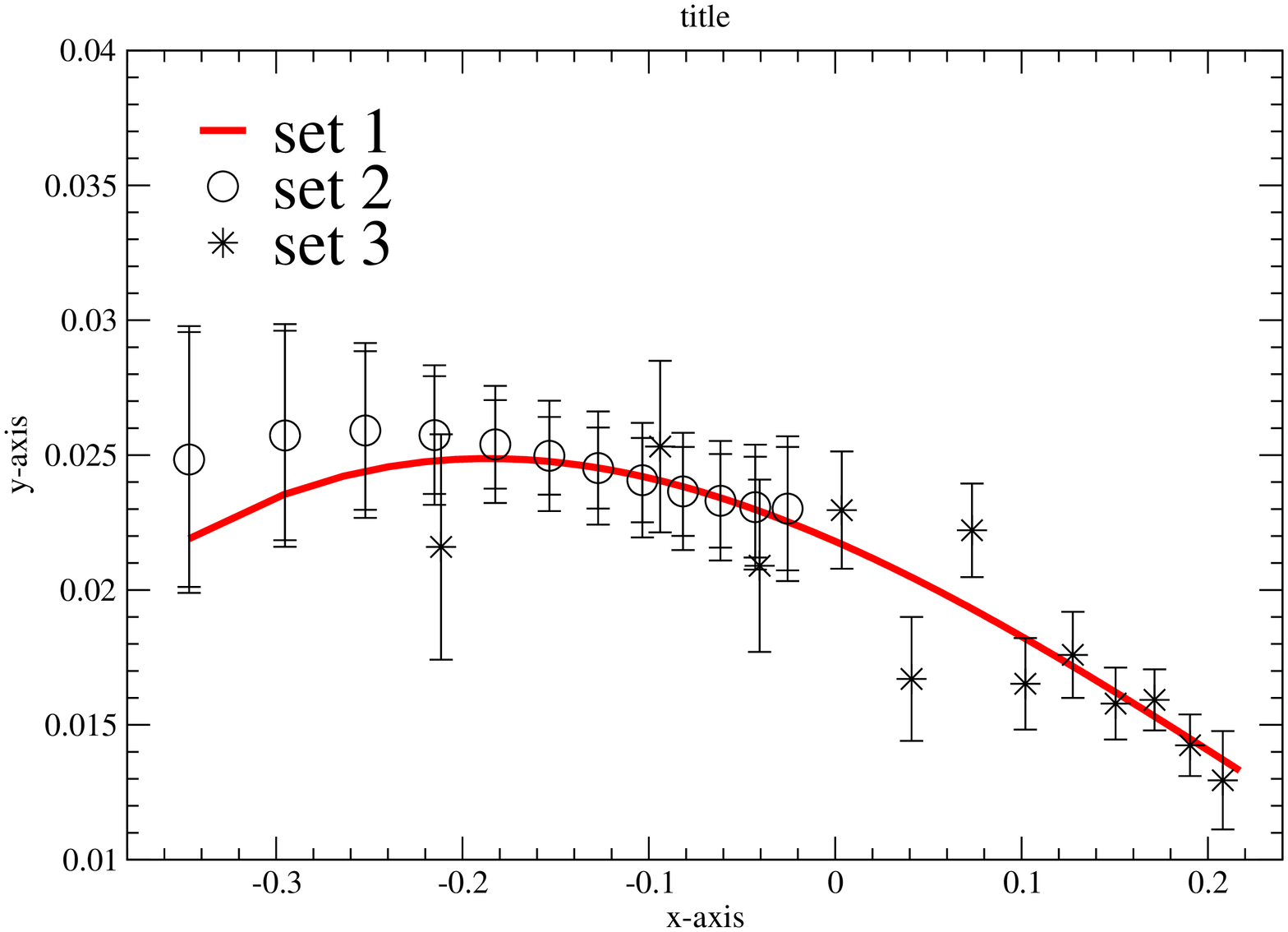}}
\end{minipage}
\end{center}
\caption{The two figures are taken from \cite{RuthBtopi}. The one in 
the left hand side shows the chiral and continuum extrapolation for 
$f_{\perp}$ which dominates the value of $f_+$. The second figure shows 
the combined z-fit of lattice and BaBar data  from which the 
result for $\vert V_{ub}\vert$ in \cite{RuthBtopi} is obtained. 
\label{BtopiRuth}}
\end{figure}

This is the approach that is being followed by the FNAL/MILC collaboration 
\cite{RuthBtopi}. The authors in \cite{RuthBtopi} 
used the so called $z-$expansion, which is a model independent 
parametrization based only on unitarity and analyticity \cite{zexp}. 
Analyticity, unitarity and heavy quark symmetry can be combined together 
with experimental data to determine the shape of the form factors.  
Lattice simulations must then only provide 
a normalization that can be extracted 
from the region where lattice data are most precise. 
The FNAL/MILC collaboration has calculated the form factors 
for two different values of the 
lattice spacing and for full QCD points. The results are extrapolated 
to the continuum and physical masses using SChPT. 
In the case of the form factor $f_{\perp}$, which dominates 
the value of $f_{+}$ defined above, 
a NLO description is enough to have good fits 
since this form factor is dominated by the $B^*$ pole. Results from the 
extrapolation are shown in Figure \ref{BtopiRuth}. 
The extrapolated lattice data are then fitted together with the experimental 
BaBar measurements \cite{babarVub} using the $z-$expansion mentioned before 
and leaving $\vert V_{ub}\vert$ as one of the free parameters 
to be determined by the fit. The output of this  analysis, 
which is plotted in Figure \ref{BtopiRuth}, is $\vert V_{ub}\vert 
= (3.38\pm 0.35)\times 10^{-3}$ \cite{RuthBtopi}. 
This value is consistent with the global fits of the CKM matrix 
\cite{ckmfits}, although $1-2\sigma$ lower than most inclusive determinations, 
which tipical values around $4-4.5\times 10^{-3}$ \cite{HFAG08}. 
In this analysis the statistical error and the one associated to the 
chiral and continuum extrapolations, which are interrelated using this method 
and thus treated as a single source of error, are dominating 
the 10\% total error in $\vert V_{ub}\vert$. Therefore, the first 
improvements that should be addressed in the future are increasing the 
statistics and doing simulations at smaller lattice spacings. 

The authors in \cite{LellouchVub08} have recently proposed a new model 
independent parametrization, which satisfies unitarity, analyticity and 
perturbative scaling. As an illustration of their method, 
they fit their parametrization of the form factor $f_+$ for $B\to\pi l\nu$ 
to experimental, existing lattice and light cone sum rules results. 
The form factors obtained yield to a value of $\vert V_{ub}\vert$ 
that agrees with the one by the FNAL/MILC collaboration.

The QCDSF collaboration is performing a quenched systematic study of 
several semileptonic decays \cite{QCDSF08}. 
They use simulations with  
$\order(a)$ improved Wilson fermions at a single lattice spacing, $a=0.04~fm$. 
The small value of $a$ allows them to simulate at the physical charm quark 
mass and bottom masses very close to the physical ones. But the light 
masses are quite heavy, $m_\pi^{min}=526{\rm MeV}$. The authors use the 
Becirevic-Kaidalov (BK) parametrization \cite{BKparametrization} 
to describe the dependence on the momentum transfer, introducing a model 
dependency in their analysis. The preliminary numbers for the form factors are 
$f^{B\to\pi}_+(0)=0.232(23)$, $f^{B\to K}_+(0)=0.29(3)$,  
$f^{D\to \pi}_+(0)=0.668(38)$, $f^{D\to K}_+(0)=0.733(38)$ and 
$f^{D_s\to K}_+(0)=0.598(20)$; where errors include both statistic and 
systematic uncertainties other than quenching and the one associated to 
the model use in the $q^2$ extrapolation.

The ratio of form factors $f^{B\to K}_+(0)/f^{B\to \pi}_+(0)$ is needed 
as an input in the prediction of the SM correlation between 
$S_{\pi^0K_S}\equiv (\sin{2\beta})_{\pi^0K_S}$ and $A_{\pi^0K_S}$ , the 
mixing-induced and direct CP asymmetries of $B^0\to \pi^0K_S$ \cite{FJPZ08}. 
The current SM prediction, which assumes the ratio to be equal to 1, is not 
in agreement with experiment, so it would be very important to study this 
channel to have an accurate determination of this ratio including sea 
quark effects in a realistic way.

\subsection{$D\to \pi(K) l\nu$: determination of  $\vert V_{cd(cs)}\vert$}

Good quantities for testing lattice QCD are the ratios of semileptonic 
and leptonic decay widths 
\ba
\frac{1}{\Gamma(D^+\to l\nu)}
\frac{d\Gamma(D\to \pi l\nu)(q^2)}{dq^2}\quad\quad
\frac{1}{\Gamma(D_s\to l\nu)}\frac{d\Gamma(D\to K l\nu)(q^2)}{dq^2}\,.
\ea
Or, equivalently, ratios of semileptonic form factors and decay constants. 
An advantage of using these ratios is that the chiral extrapolation 
to the physical pion mass is milder than for denominator or numerator alone. 

On the other hand, these $D$ semileptonic decays can be used to extract 
the CKM matrix elements $\vert V_{cs}\vert$ and $\vert V_{cd}\vert$. 
In fact, the experimental measurement of the branching ratio 
$Br\left(D\to K e\nu\right)$ and the lattice calculation of the corresponding 
form factors by the FNAL/MILC collaboration \cite{fnalDtoK}  
are the inputs of the current best determination of 
$\vert V_{cs}\vert=1.015\pm0.015\pm0.106$. The 
semileptonic decay $D\to \pi e\nu$ has also the potential to provide 
the most accurate result for $\vert V_{cd}\vert$ if errors in the 
lattice calculation of the form factors are reduced 
to the 10\% level or less.

\begin{figure}
\begin{center}
\includegraphics[width=0.5\textwidth]{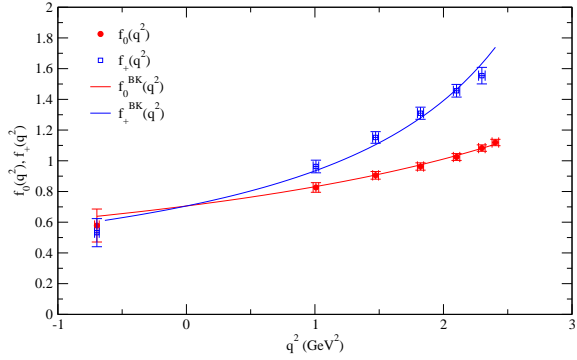}
\caption{Data points for the form factors $f_+^{D\to\pi l\nu}(q^2)$ 
and $f_0^{D\to\pi l\nu}(q^2)$ 
for different values of $q^2$ together with the corresponding BK 
parametrizations $f_{+(0)}^{BK}(q^2)$. 
Figure courtesy of S.~Simula.\label{ETMCDtopi}}
\end{center}
\end{figure}

The ETMC collaboration is calculating the form factors for $D\to\pi(K) l\nu$ 
using $N_f=2$ twisted mass QCD at maximal twist. The main new techniques 
applied in this calculation are the all-to-all propagators obtained with 
a stochastic method and twisted boundary conditions. The 
main limitations of the preliminary results in \cite{ETMDtopi} 
are the fact that the simulations are performed at a single 
lattice spacing and the use of a BK parametrization to describe the $q^2$ 
dependence. The preliminary results obtained for $D\to\pi l\nu$ as a 
function of $q^2$ can be seen in Figure \ref{ETMCDtopi}. 
The physical pion masses are achieved by an extrapolation of the data 
for the ratio of the form factor $f_+(q^2)$ and the decay constant $f_{D}$ 
with $0.3\le m_\pi({\rm GeV})\le 0.6$. They reached the physical 
$D$ mass by an interpolation of results at four different heavy 
quark masses around $m_c$.

\begin{figure}
\begin{center}
\includegraphics[width=0.5\textwidth]{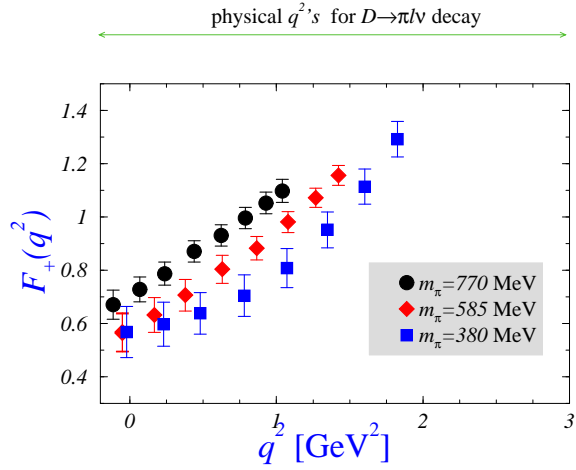}
\caption{Results for the form factor $f_+^{D\to\pi l\nu}(q^2)$ 
from simulations using three different values of the light quark masses. 
Figure courtesy of B.~Haas.\label{fig:Haas08}}
\end{center}
\end{figure}

Becirevic, Haas and Mescia also presented preliminary results of 
their calculation of the form factors for $D\to\pi l\nu$ with 
$N_f=2$ flavours of sea quarks, employing the configurations by the 
QCDSF collaboration in \cite{Haas08}. They use $\order(a)$ improved Wilson 
fermions at a single lattice spacing $a\simeq0.08~fm$, pion masses 
$m_\pi=770,585,380~{\rm MeV}$ and a fixed value of the charm valence  
quark mass around the physical one. The method 
employed is the double ratio strategy 1 described in \cite{BHM07}, where 
the $D$ meson is at rest and the momentum is injected to the pion. 
Twisted boundary conditions are also used in this work. 
The preliminary results obtained for $f_+$ are collected in Figure 
\ref{fig:Haas08} for the different light quark masses simulated. 
In this Figure, one can appreciate a qualitative change in the shape 
of the the form factor, that also occurs for $f_0$, when $m_\pi$ goes 
to the physical value. That corresponds to the better resolution of 
the polar behaviour due to $m_{D^*}$ when the data are moving closer to 
the physical kinematic region and point out the importance of having 
light masses small enough. The chiral extrapolation in this work is 
also done for the ratio of the semileptonic form factor and the decay 
constant. The authors performed both a extrapolation using HMChPT and 
a linear extrapolation, since they didn't observe any sensitivity 
of their data to the logarithms. The difference between the 
results coming from the two fits will be added as a source of 
systematic error.

The results from both works, together with the corresponding experimental 
number obtained from CLEO-c results in \cite{cleoc08} and 
\cite{cleocformfactor} (using the z-expansion parametrization) are 
in Table \ref{DtoKresults}.
\begin{table}
\begin{center}
\begin{tabular}{cc}
\hline
Reference & $[f_+^{D\to\pi}(q^2=1{\rm GeV}^2)/f_{D^+}]{\rm GeV}^{-1}$\\
\hline
ETMC \cite{ETMDtopi} & $4.39\pm0.31_{stat.}$\\
Becirevic \emph{et al} (linear fit) \cite{Haas08} & $3.76\pm0.54$\\
Becirevic \emph{et al} (HMChPT fit) \cite{Haas08} & $4.32\pm0.56$\\
CLEO \cite{cleoc08,cleocformfactor} & $4.51\pm0.53$\\
\hline
\end{tabular}
\end{center}
\caption{Results for  $[f_+^{D\to\pi}(q^2=1{\rm GeV}^2)/f_{D^+}]
{\rm GeV}^{-1}$ from the $N_f=2$ calculations described in the text and 
CLEO experimental results. \label{DtoKresults}}
\end{table}
Both analyses need to have a better control over the extrapolation to 
the physical masses. They also need to explicitly study discretization 
errors with simulations at several lattice spacings.

\section{$B^0-\bar B^0$ mixing}

\label{B0mixing}

The mixing in the $B^0_q-\bar B^0_q$ system is an interesting  
place to look for NP effects. The BSM effects can appear 
as new tree level contributions, or through the presence of new particles 
in the box diagrams. In fact, it has been recently claimed that there is 
a disagreement between the direct experimental 
measurement of the phase of $B^0_s$ mixing 
amplitude and the SM prediction \cite{UTfit08}. Possible NP effects have also 
been reported to show up in the comparison between direct experimental 
measurements of $\sin(2\beta)$ and SM predictions using $B^0$ 
mixing parameters \cite{LunguiSoni08}. Studies of neutral $B$ meson 
mixing parameters can also impose important constraints on different 
NP scenarios \cite{fleischer}. 

In the SM, the $B^0-\bar B^0$ mixing is due to box diagrams with 
exchange of two $W$-bosons. These box diagrams can be rewritten in terms of 
an effective Hamiltonian with four-fermion operators describing 
processes with $\Delta B=2$. The matrix elements of the operators 
between the neutral meson and antimeson encode the non-perturbative 
information on the mixing and can be calculated using lattice QCD 
techniques. Those matrix elements  are parametrized by products 
of $B$ decay constants and bag parameters, which 
provide the value of quantities experimentally measurable, like the 
mass differences,  $\Delta M_{s,d}$, and decay width differences, 
$\Delta \Gamma_{s,d}$, between the heavy and light $B^0_s$ and 
$B^0_d$ mass eigenstates. For example, the mass difference 
is given by
\be\label{SMMsd}
\Delta M_{s(d)}\vert_{theor.}\propto
\vert V_{t s(d)}^*V_{tb}\vert^2
f_{B_{s(d)}}^2\hat B_{B_{s(d)}}\, ,
\ee
with  $\langle \bar B^0_s\vert Q^{s(d)}_L \vert B^0_s
\rangle = \frac{8}{3}M^2_{B_{s(d)}}f^2_{B_{s(d)}}
B_{B_{s(d)}}(\mu)$ and $O^{s(d)}_L = \left[\bar b^i 
\gamma_\mu(1-\gamma_5)s^i(d^i)\right]\,
\left[\bar b^j\gamma^\mu(1-\gamma_5)s^j(d^j)\right]$.

Many of the uncertainties that affect the theoretical calculation
of the decay constants and bag parameters cancel totally or partially 
if one takes the ratio $\xi^2=f_{B_s}^2 B_{B_s}/f_{B_d}^2 B_{B_d}$. 
Hence, this ratio and therefore the combination of CKM matrix elements 
related to it, $\vert \frac{V_{td}}{V_{ts}}\vert$,  
can be determined with a significantly 
smaller error than the individual matrix elements. The ratio $\xi$ is also 
an important ingredient in the unitarity triangle analyses and the 
search for BSM effects \cite{LunguiSoni08}.

The first lattice calculation of the $B^0$ mixing parameters with 
$N_f=2+1$ sea quarks, which only studied the $B^0_s$ sector, was 
performed by the HPQCD collaboration in \cite{ourbsmix}. The authors 
obtained 
\be\label{masaresult}
\Delta M_s = 20.3(3.0)(0.8)ps^{-1}\quad{\rm and}\quad
\Delta \Gamma_s = 0.10(3)ps^{-1}\,, 
\ee
which is compatible with experiment. 

The FNAL/MILC \cite{B0mixingfnal} and HPQCD \cite{B0mixinghpqcd} 
collaborations are currently working on a more 
complete study of $B^0-\bar B^0$ mixing, including $B^0_s$ and 
$B^0_d$ parameters. The main goal of both projects is obtaining the ratio 
$\xi$ fully incorporating vacuum polarization effects. 
The choice of actions and the setup is the same as for their $f_B$, 
$f_{B_s}$ calculations described in Section \ref{sec:fbs}.

\begin{figure}[h]
\begin{minipage}[c]{0.44\textwidth}
\begin{center}
\includegraphics[width=0.7\textwidth,angle=-90]{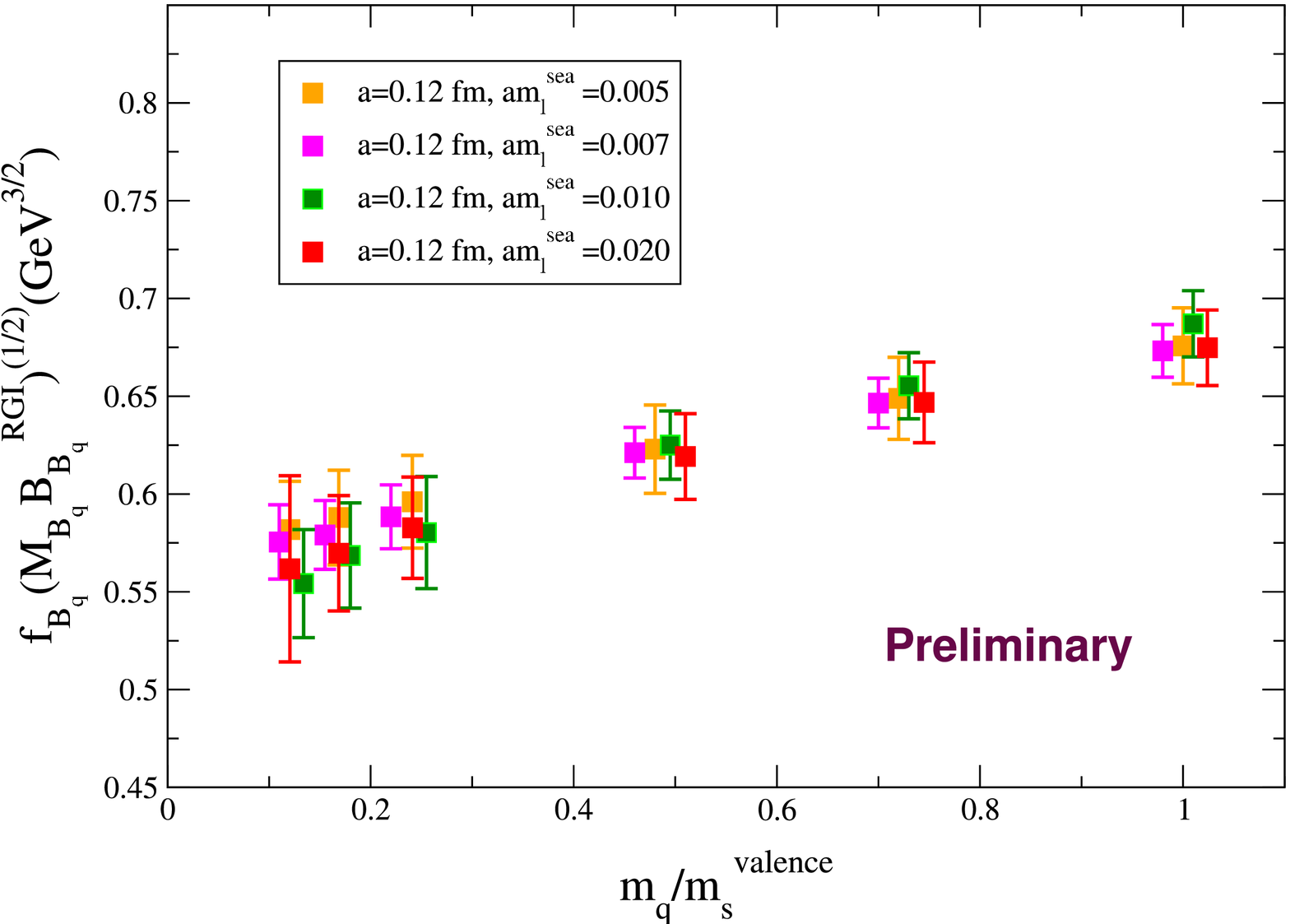}
\end{center}
\caption{Bare values of $f_{B_q}\sqrt{M_{B_q} B_{B_q}}$ in lattice units 
as a function of the light valence quark mass $m_q$ normalized to the value 
of the physical strange quark mass from the FNAL/MILC collaboration. 
The results correspond  
to one of the three lattice spacings at which the  FNAL/MILC's study 
is performed. The bottom valence quark is fixed 
to its physical value and the strange valence quark is very close 
to its physical value. The strange sea quark mass is also very close to its 
physical value.
\label{fBBBFNAL}}
\end{minipage}
\hspace*{0.8cm}\begin{minipage}[c]{0.44\textwidth}
\begin{center}
\includegraphics[width=0.7\textwidth,angle=-90]
{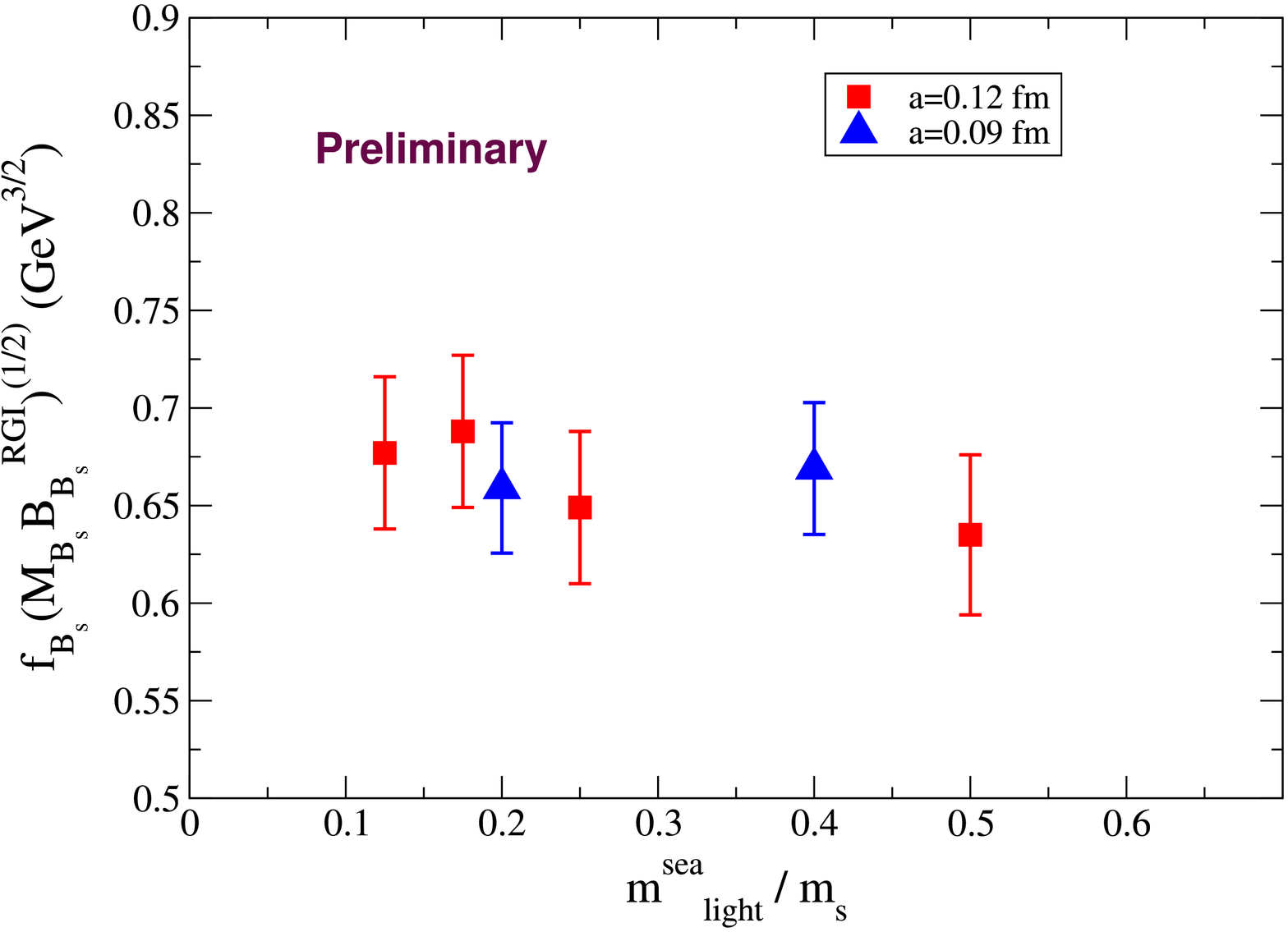}
\end{center}
\caption{Values of $f_{B_s}\sqrt{M_{B_s}\hat B_{B_s}}$ in ${\rm GeV}^{3/2}$ 
as a function 
of the light sea quark mass normalized to the physical strange quark mass 
from the HPQCD collaboration. The data include statistical, 
perturbative and scale errors. The bottom valence quark is fixed 
to its physical value and the strange valence quark is very close 
to its physical value. The strange sea quark mass is also very close to its 
physical value.\label{fBBBHPQCD}}
\vspace*{1.0cm}
\end{minipage}
\end{figure}

Figures  \ref{fBBBFNAL} and \ref{fBBBHPQCD} show some examples of 
the preliminary values of $f_{B_q}\sqrt{M_{B_q} B_{B_q}}$ obtained 
as function of light valence mass and light sea quark masses respectively. 
The renormalization of the matrix elements is done in both cases 
perturbatively at one-loop. However, the FNAL/MILC collaboration 
results for the matching coefficients are still preliminary, 
so only bare results from this collaboration are shown 
in Figure \ref{fBBBFNAL}.

The dependency on the light sea quark mass is in both studies very mild 
as can be seen in the Figures, so only the chiral extrapolation in the 
$d$ quark mass for $B^0_d$ parameters is expected to be a significant 
source of error. Figure \ref{fBBBHPQCD} also shows 
that the results for the two different lattice spacings 
are very similar, which indicates small discretization errors. The 
statistical errors in both cases and for parameters in both the $B^0_s$ 
and $B^0_d$ systems are in the range 1-4\%. Relativistic corrections after 
power law subtractions are under control in the HPQCD study. They are 
around 5-6\% for the coarse lattice and 3-4\% for the fine lattice.

\begin{figure}[h]
\begin{center}
\includegraphics[width=0.40\textwidth,angle=-90]{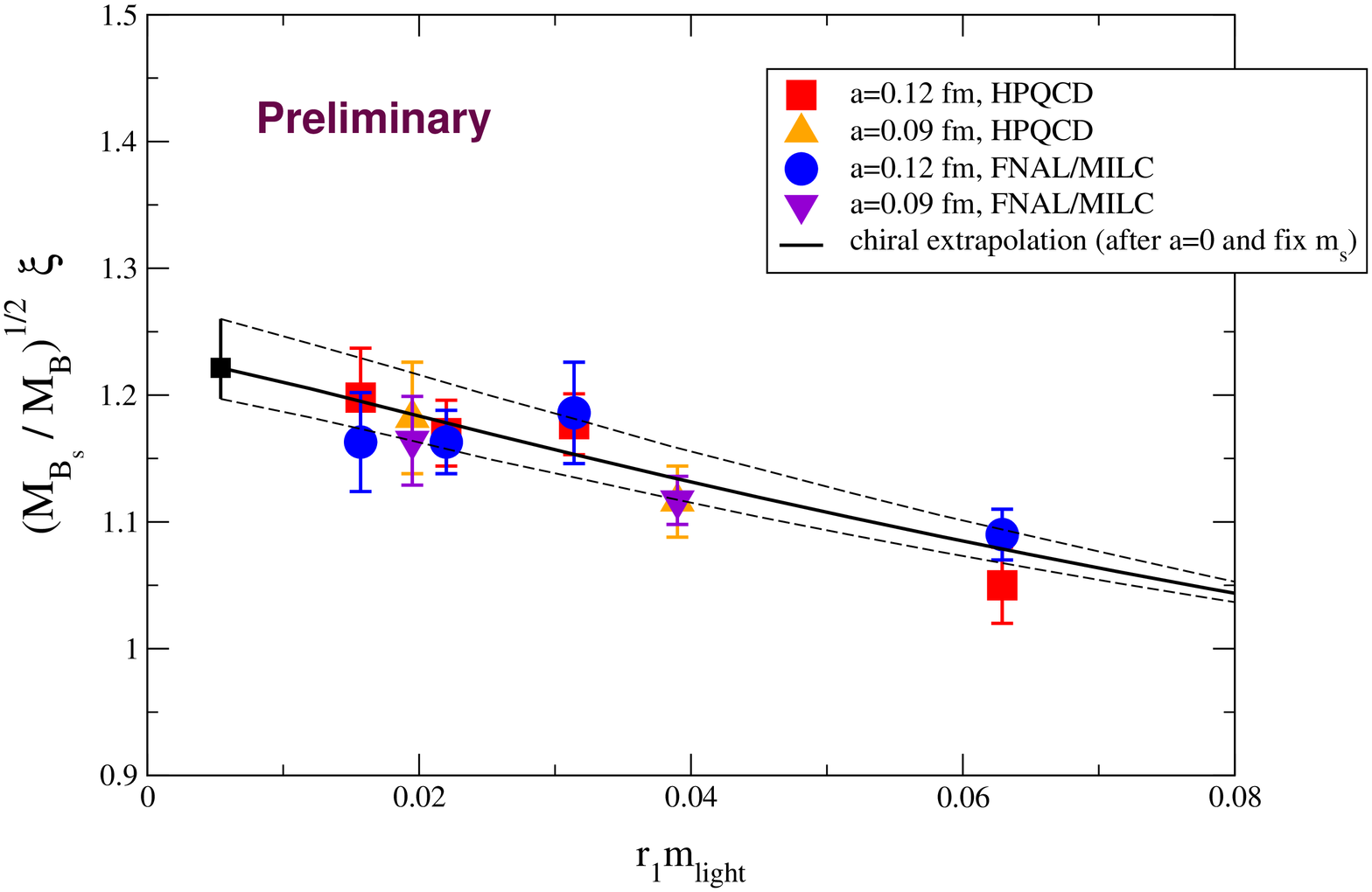}
\end{center}
\caption{Product of the ratios $\xi$ and $M_{B_s}/M_{B_d}$ as 
a function of the down quark mass normalized to the strange quark mass. 
Results for both FNAL/MILC and HPQCD collaborations including only 
statistical errors are shown for two different values of the lattice 
spacing $a$. Only the full QCD points are shown for the FNAL/MILC 
collaboration. \label{xicomparison}} 
\end{figure}

A comparison of the preliminary results from the two collaborations 
for the ratio $\xi$ is shown in Figure \ref{xicomparison}.
Again, FNAL/MILC results are non-renormalized ones, but in the case of $\xi$ 
a strong cancellation of perturbative corrections is expected between 
numerator and denominator, so renormalized results are going to be shifted 
by less than a 1\% with respect to the bare ones. 
This has been explicitly checked in  
the HPQCD analysis and in the FNAL/MILC analysis with the preliminary 
matching coefficients. The differences between 
fine and coarse lattices in Figure \ref{xicomparison} are small. This  
suggests small discretization effects. Another observation about 
that figure is that there is a very good agreement between 
the results from the two collaborations. Hence, heavy quark discretization 
effects, which are different for the two collaborations, must be small.

Figure \ref{xicomparison} also contains the curve corresponding to the 
extrapolation to the physical $m_d$ after $m_s$ is fixed to its 
physical value and the continuum limit is taken, from the 
FNAL/MILC collaboration. That is done using SChPT at NLO plus analytic 
NNLO terms. The preliminary result from this collaboration after 
extrapolations are performed is $\xi=1.211\pm0.038\pm0.024$, where the 
second error is the sum in quadrature of the systematic uncertainties 
some of which are still under investigation.

The ALPHA collaboration has completed the computation of the 
non-perturbative renormalization and renormalization group running 
of the complete basis of four fermion operators with $\Delta B=2$ 
using $N_f=2$ dynamical Wilson fermions \cite{AlphaF2}. Heavy 
quarks are described in this analysis in the static limit. Using 
Wilson actions with suitable twisted mass terms, the mixing parameters 
can be related to matrix elements of parity-odd operators that 
are protected from extra mixings under renormalization due to the 
breaking of chiral symmetry. The precision of the final study of 
$B^0-\bar B^0$ parameters using these results is however limited 
due mainly to cut-off effects. That could be improved in the future 
by simulating at smaller values of the lattice spacing and/or 
improving operators at order $a$.

The RBC/UKQCD collaboration presented last year at this conference 
preliminary  $N_f=2+1$ bare results for the $B^0-\bar B^0$ parameters 
\cite{WennekersLat08}. This year the collaboration has discussed in this 
conference the values of the renormalization coefficients calculated to 
one-loop \cite{tomomiLat08}. The renormalization could partially 
correct the differences found in \cite{WennekersLat08} between the results 
using different smearings.

\subsection{$B^0-\bar B^0$ mixing beyond the Standard Model}

The effects of heavy new particles in the box diagrams that describe 
the $B^0$ mixing can be seen in the form of effective 
operators built with SM degrees of freedom. The NP could modify the 
Wilson coefficients of the four-fermion operators that already contribute 
to $B^0$ mixing in the SM and gives rise to new four-fermion operators in 
the $\Delta B=2$ effective Hamiltonian -see \cite{gabbiani,damir} for a list 
of the possible operators in the SUSY basis. The calculation of 
those Wilson coefficients for a particular BSM theory, together with 
the lattice calculation of the matrix elements of all the possible 
four-fermion operators in the SM and beyond and experimental 
measurements of $B^0$ mixing parameters, can constrain BSM 
parameters and help to understand NP. 

To date, there does not 
exist an unquenched determination of the complete set of matrix elements 
of four-fermion operators in that general $\Delta B=2$ effective Hamiltonian. 
However, both FNAL/MILC and HPQCD collaborations are working on  
extending their analysis to BSM operators in the near 
future. Actually, the HPQCD collaboration has already calculated 
the one-loop matching coefficients needed for such an analysis 
\cite{Bb_bsm}.

\section{Heavy quark masses}

\subsection{Charm quark mass}

A new method to extract the charm quark mass has been proposed this year 
in \cite{mcHPQCD08}. The method is analogous to the extraction of 
$m_c$ from dispersion relations using perturbative determination of 
zero-momentum moments of current-current correlators and experimental 
data from $e^+e^-\to hadrons$ \cite{mccontinuum}. In \cite{mcHPQCD08}, 
experimental data are substituted by lattice data. The methodology 
proposed in \cite{mcHPQCD08} is interesting in several ways. First of all, 
it provides a way of calculating $m_c$ that does not rely on 
lattice perturbation theory but continuum perturbation theory, for 
which higher orders in the expansion are known. The precision achieved 
is thus better than with the traditional methods. It also constitutes a 
way of checking the lattice discretization and techniques used in 
the calculation, since the result can be compared against the one 
coming from the pure continuum calculation. That would give confidence 
to apply the same discretization formalism and techniques to 
other charm quantities, like decay constants. 
Finally, the same methodology can be applied to 
different correlators to extract other masses, condensates, etc. 

\begin{figure}[th]
\begin{minipage}[c]{0.44\textwidth}
\begin{center}
{\includegraphics[width=1.\textwidth]{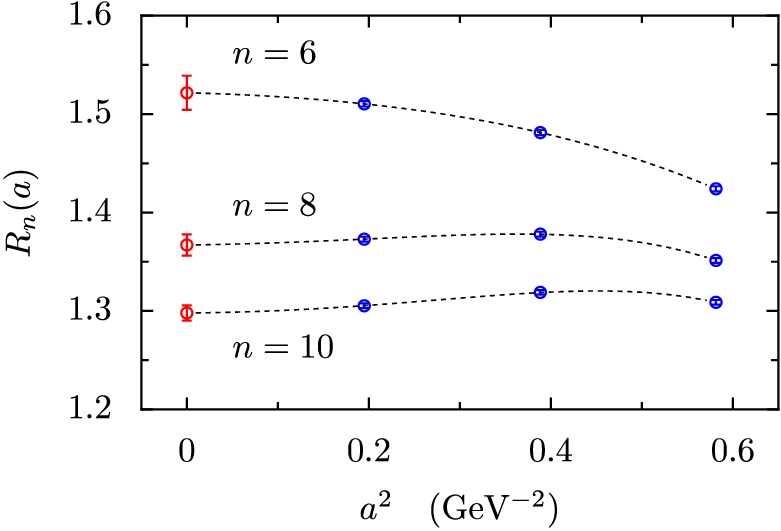}}
\caption{Extrapolation to the continuum for some of the moments 
$R_n$ analyzed to get the final value of $m_c$. 
Figure from \cite{mcHPQCD08}. 
\label{Rnadependence}}
\end{center}
\end{minipage}
\hspace*{0.8cm}\begin{minipage}[c]{0.44\textwidth}
\begin{center}
{\includegraphics[width=1.\textwidth]{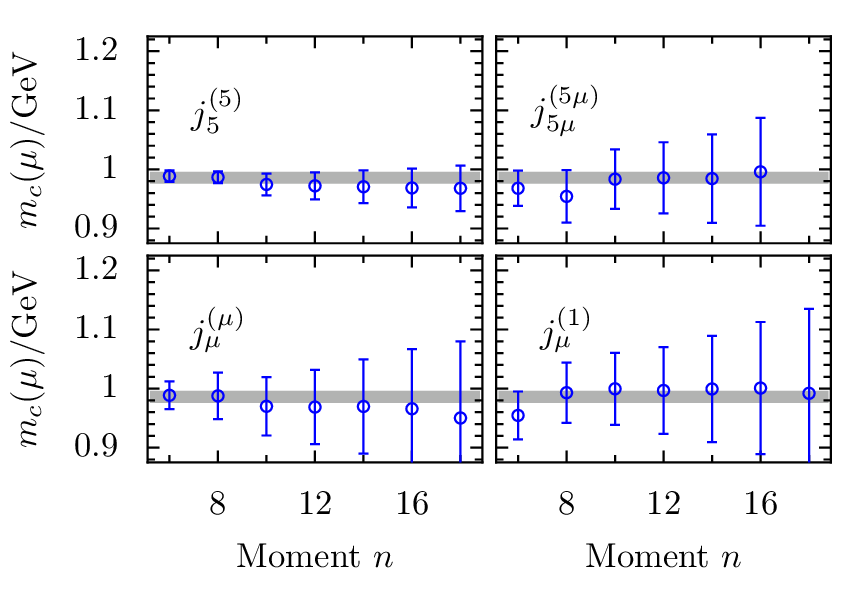}}
\caption{Values of $m_c(3{\rm GeV})$ from different moments of 
correlators and different lattice operators. The grey band is the 
final result for the mass. Figure from \cite{PeterLat08}. 
\label{correlators}}
\end{center}
\end{minipage}
\end{figure}

The charm quark mass is extracted in \cite{mcHPQCD08} from moments of 
charm quark correlators calculated with the HISQ action on $N_f=2+1$ 
MILC configurations. The moments are defined as 
$G_n \equiv \sum_t (t/a)^nG(t)$, where the correlation function $G(t)$ 
is given by 
\ba\label{correlatormc} 
G(t)\equiv a^6\sum_{\vec{x}}(am_{0c})^2\langle 0 
\vert j_5(\vec{x},t)j_5(0,0)\vert 0\rangle\, ,
\ea 
with $j_5=\bar \psi_c \gamma_5 \psi_c$. On the other hand, those 
moments can be calculated in the continuum through the expression 
$G_n = \frac{g_n(\alpha_{\overline{MS}}(\mu),\mu/m_c)}
{\left(am_c^{\overline{MS}}(\mu)\right)^{n-4}}$, where 
$g_n(\alpha_{\overline{MS}}(\mu),\mu/m_c)$ is evaluated perturbatively 
and for some values of $n$ it is known to four loops. 
The normalization factors in (\ref{correlatormc}) are such that 
the lattice and continuum expressions can be matched in the continuum 
limit, allowing the extraction of $m_c^{\overline{MS}}$ from that matching 
relation. Appropriate ratios of the moments, $R_n$, are defined so systematic 
uncertainties like the $\order\left((am_c)^n\right)$ errors and the one 
associated with the tuning of $am_{0c}$ and the lattice spacing are 
suppressed. The resulting dependence on the lattice spacing is very mild, 
in particular for moments with $n\ge 8$, as can be seen in Figure 
\ref{Rnadependence}.   
The authors in \cite{mcHPQCD08} tested the impact of systematic 
errors and taste-changing effects due to the use of staggered fermions 
by calculating $m_c$ with different moments and different vector 
and axial correlators. Lattice artifacts enter in a different way in those 
calculations, so the agreement between them, shown in Figure 
\ref{correlators}, indicates that those artifacts are very 
much under control in this study.

Updated results from this analysis was presented at this conference 
\cite{PeterLat08} and later appeared in \cite{mcHPQCD08}. 
The updates include new simulations at a fourth, smaller, lattice 
spacing ($a=0.06~fm$) and higher 
order terms in the continuum perturbative expansion for some of the moments. 
The updated value of the charm quark mass is $m_c^{\overline{MS}}(3{\rm GeV}) 
= 0.986(10){\rm GeV}$. This agrees perfectly with the one obtained from 
the purely continuum calculation using $e^+e^-\to hadrons$ data, 
$m_c^{\overline{MS}}(3{\rm GeV}) = 0.986(13){\rm GeV}$. 

The same methodology can be applied to the calculation of the bottom 
quark mass. This is being done by the same collaboration and 
preliminary results were also reported in \cite{PeterLat08}. 
They used the NRQCD action for the $b$ quarks. In this case, currents 
are not conserved and extra ratios of moments must be taken in order 
to cancel the renormalization factors.  
Their result is $m_b^{\overline{MS}}(m_b) = 4.20(4){\rm GeV}$, 
that again agrees very well with the continuum one, 
$m_b^{\overline{MS}}(m_b) = 4.16(3){\rm GeV}$. 

The HPQCD collaboration is also calculating $m_c$ with HISQ fermions 
on $N_f=2+1$ MILC configurations in a more standard way, 
using two-loop perturbation theory based 
on traditional methods and high-$\beta$ techniques. The preliminary 
value, $m_c^{\overline{MS}}(3{\rm GeV}) = 0.983(25){\rm GeV}$,  
\cite{mcAllison} agrees with the one obtained from 
the current-current correlators method, but with significantly 
larger errors. The main goal of this calculation is 
obtaining an accurate determination of $m_c/m_s$ that, together with 
a precise value of $m_c$, will allow getting a value of $m_s$ with 
a very small error. It will be interesting to compare this value of 
$m_s$ with the one obtained in a direct way by the same collaboration 
with Asqtad fermions in \cite{msHPQCD}.

Another on-going $N_f=2+1$ determination of $m_c$ and $m_b$ is the one 
by the FNAL/MILC collaboration, whose preliminary results were 
presented at this conference last year \cite{ElizLat07}. The main 
error there is the one associated with the truncation of the 
perturbative series.

\subsection{Bottom quark mass}

An accurate determination of the bottom quark mass is crucial for 
many continuum perturbation theory studies of heavy flavour observables. 
For example, the inclusive determination of $\vert V_{ub}\vert$ is 
very sensitive to the value of $m_b$ used in the theoretical analysis.

Besides the preliminary results presented for $m_b$ by the HPQCD 
collaboration using the current-current correlators method and 
mentioned in the last section, only quenched determinations of this 
parameter have appeared recently.

The ALPHA collaboration calculated this parameter in the framework of 
HQET including $1/m$ corrections in \cite{alphamb}. The result is 
$m_b^{\overline{MS}}(m_b) = 4.347(48){\rm GeV}$, very much compatible with 
the number quoted in \cite{GST08}, $m_b^{\overline{MS}}(m_b) 
= 4.42(6){\rm GeV}$. 
That second result is also obtained in the quenched approximation and 
in the framework of HQET going beyond the static limit, but it is 
based on the SS method. The setup is the same as that discussed for the 
$f_{B_s}$  in section \ref{sec:fbs}. Again, it is important that these 
two different approaches agree, but an unquenched calculation is needed 
to have reliable results.

A completely different approach is followed by the TWQCD collaboration 
in \cite{TWQCDmass}. That reference describes an exploratory study in 
the quenched approximation based on the use of relativistic domain wall 
fermions simulated in a very small volume (very small lattice spacing). 
They got $m_c^{\overline{MS}}(m_c)=1.16\pm0.04{\rm GeV}$ and 
$m_b^{\overline{MS}}(m_b)=4.65\pm0.05{\rm GeV}$, where the errors are 
estimates which do not include all the systematics. In \cite{TWQCDmass}, 
the authors predicted a mass of the $\eta_b$, $m_{\eta_b}=9383(4)(2){\rm MeV}$ 
which agrees with the subsequent experimental measurement by the BaBar 
collaboration \cite{babar_etab}, $m_{\eta_b}=9388.9^{+3.1}_{-2.3}
\pm2.7{\rm MeV}$.

\section{Conclusions}

The non-perturbative input coming from lattice calculations 
of hadronic matrix elements in the heavy flavour 
sector is needed to extract 
Standard Model parameters and test the model itself by overconstraining 
the values of those parameters. The information encoded in such  
heavy flavour observables is complementary to direct searches in LHC 
to study and constrain New Physics effects. 
Hints of discrepancies between SM predictions and experimental 
measurements have started to show up in some CP violating 
observables \cite{BG08,LunguiSoni08} in the $B^0-\bar B^0$ system 
and charm meson leptonic decays \cite{BK08}. The precise determination of the 
CKM matrix elements $\vert V_{ub}\vert$ and $\vert V_{cb}\vert$, 
and the accurate 
calculation of parameters like $\xi$, decay constants, $\dots$, involved 
in those analyses, is crucial in order to discern the origin of those 
disagreements and fully exploit the potential of the CP violating 
observables on constraining NP. Those calculations must be prioritized.

In order to be able to get the around 5\% accuracy required by 
phenomenology, vacuum polarization effects must be included in the 
simulations in a realistic way, i.e., $N_f=2+1$,  and a rigorous 
study of systematic errors, 
including an analysis of the validity of the ChPT techniques used,  
must be performed. Such calculations are now possible. The use of 
techniques like double ratio methods, twisted boundary conditions and 
model independent parametrizations of the $q^2$ dependence of form 
factors have allowed (and/or will allow) to make important 
progress in achieving those goals. 

New precise results have been presented at this conference for the $D$ and 
$B$ decay constants, the form factor describing $B$ semileptonic decays and 
the charm quark mass. Additional analyses of those parameters from other  
collaborations, together with final results for $B^0$ mixing parameters, $D$ 
semileptonic decay form factors, the bottom quark mass, will soon be  
available with errors at the few percent level. The extension of 
$B^0-\bar B^0$ studies in the SM to include BSM operators and to 
study short-distance contributions to $D^0-\bar D^0$ mixing in the 
SM and beyond are being pursued by the HPQCD and FNAL/MILC collaborations.

\acknowledgments

I thank Ian Allison, Benoit Blossier, Ting-Wai Chiu, Christine Davies, 
Aida El-Khadra, Todd Evans, Eduardo Follana, Benjamin Haas, Jack Laiho, 
Peter Lepage, Paul Mackenzie, Marco Panero, Silvano Simula, 
Junko Shigemitsu, Amarjit Soni, Nazario Tantalo, Cecilia Tarantino, 
Ruth Van de Water and Georg von Hippel for help in the preparation 
of this talk. I thank the organizers to invite me to this enjoyable meeting. 
This work was supported in part by the DOE 
under grant no. DE-FG02-91ER40677, by the Junta de Andaluc\'{\i}a 
[P05-FQM-437 and P06-TIC-02302] and by the URA Visiting Scholars Program.

\end{document}